\documentclass[%
 reprint,
 twocolumn,showpacs,
 amsmath,amssymb,
 aps,
 prd,showkeys,superscriptaddress
]{revtex4-2}

\usepackage{graphicx}
\usepackage{dcolumn}
\usepackage{bm}
\usepackage{float}
\usepackage{xfrac}
\usepackage{color}
\usepackage[normalem]{ulem}
\usepackage{mathtools}
\usepackage{scalerel}
\usepackage{upgreek}
\usepackage{soul}
\usepackage{mathrsfs}  
\usepackage[unicode=true,pdfusetitle,
 bookmarks=true,bookmarksnumbered=false,bookmarksopen=false,
 breaklinks=false,pdfborder={0 0 1},backref=false,colorlinks=true, citecolor=blue]
           {hyperref}
           
\begin{document}

\preprint{APS/123-QED}

\title{Bulk-Surface Event Discrimination in Point Contact Germanium Detectors at Near-Threshold Energies with Shape-Matching Pulse-Shape Methods}

\newcommand{\as}{Institute of Physics, Academia Sinica, Taipei 115201, Taiwan.}
\newcommand{\bhu}{Department of Physics, Banaras Hindu University, Varanasi 221005, India.}

\newcommand{\corrjsw}{jswang2024@gate.sinica.edu.tw}
\newcommand{\corrms}{manu@gate.sinica.edu.tw}
\newcommand{\corrlhb}{lihb@gate.sinica.edu.tw}
\newcommand{\corrhtw}{htwong@gate.sinica.edu.tw}

\author{ Jia-Shian Wang} \altaffiliation[]{ \corrjsw } \affiliation{ \as } 
\author{ Manoj~Kumar~Singh}  \altaffiliation[]{ \corrms } \affiliation{ \as }  \affiliation{ \bhu } 
\author{ Hau-Bin Li} \altaffiliation[]{ \corrlhb } \affiliation{ \as }
\author{ Henry T. Wong} \altaffiliation[] { \corrhtw }  \affiliation{ \as }

\date{\today}

\begin{abstract}

The p-type point-contact germanium ~(\textit{p}PCGe) detectors have been widely adopted in searches for low energy physics events such as neutrinos and dark matter.
This is due to their enhanced capabilities of background rejection, sensitivity at energies as low as the sub-keV range and particularly fine energy resolution.
Nonetheless, the \textit{p}PCGe is subject to irregular behaviour caused by surface effects for events near the passivated surface.
These surface events can, in general, be distinguished from events that occur in the germanium crystal bulk by its slower pulse rise time.
Unfortunately, the rise-time spectra of bulk and surface events starts to convolve with each other at sub-keV energies.
In this work, we propose a novel method based on cross-correlation shape-matching combined with a low-pass filter to constrain the initial parameter estimates of the signal pulse. 
This improvement at the lowest level leads to a 50\% reduction in computation time and refinements in the rise-time resolution, which will, in the end, enhance the overall analysis.
To evaluate the performance of the method, we simulate artificial pulses that resembles bulk and surface pulses by using a programmable pulse generator module~(pulser).
The pulser-generated pulses are then used to examine the pulse behaviours at near-threshold energies, suggesting a roughly 70\% background-leakage reduction in the bulk spectrum.
Finally, the method is tested on data collected from the TEXONO experiment, where the results are consistent with our observations in pulser and demonstrated the possibility of lowering the analysis threshold by at least 10~eV$_{ee}$.

\end{abstract}

\maketitle

\section{Introduction} 
\label{sec:intro}

The point-contact germanium~(PCGe) detector, first proposed in the 1980s~\cite{Luke1989}, is a cylindrical high-purity germanium detector design where the coaxial electrode is reduced to a point-like contact.
With this configuration, depletion capacitance on the order of $\mathcal{O}(\mathrm{pF})$ can be achieved, rendering sub-keV energy thresholds possible.
Consequently, PCGe detectors generally feature the properties of fine energy resolution at sub-keV energies and low electronic noise level, making them ideal for low energy physics searches.
PCGes have, therefore, been widely and successfully employed in experiments targeting on rare physics processes such as low-energy neutrinos or cold dark matter.

\begin{figure}
    \centering
    \includegraphics[width=\columnwidth]{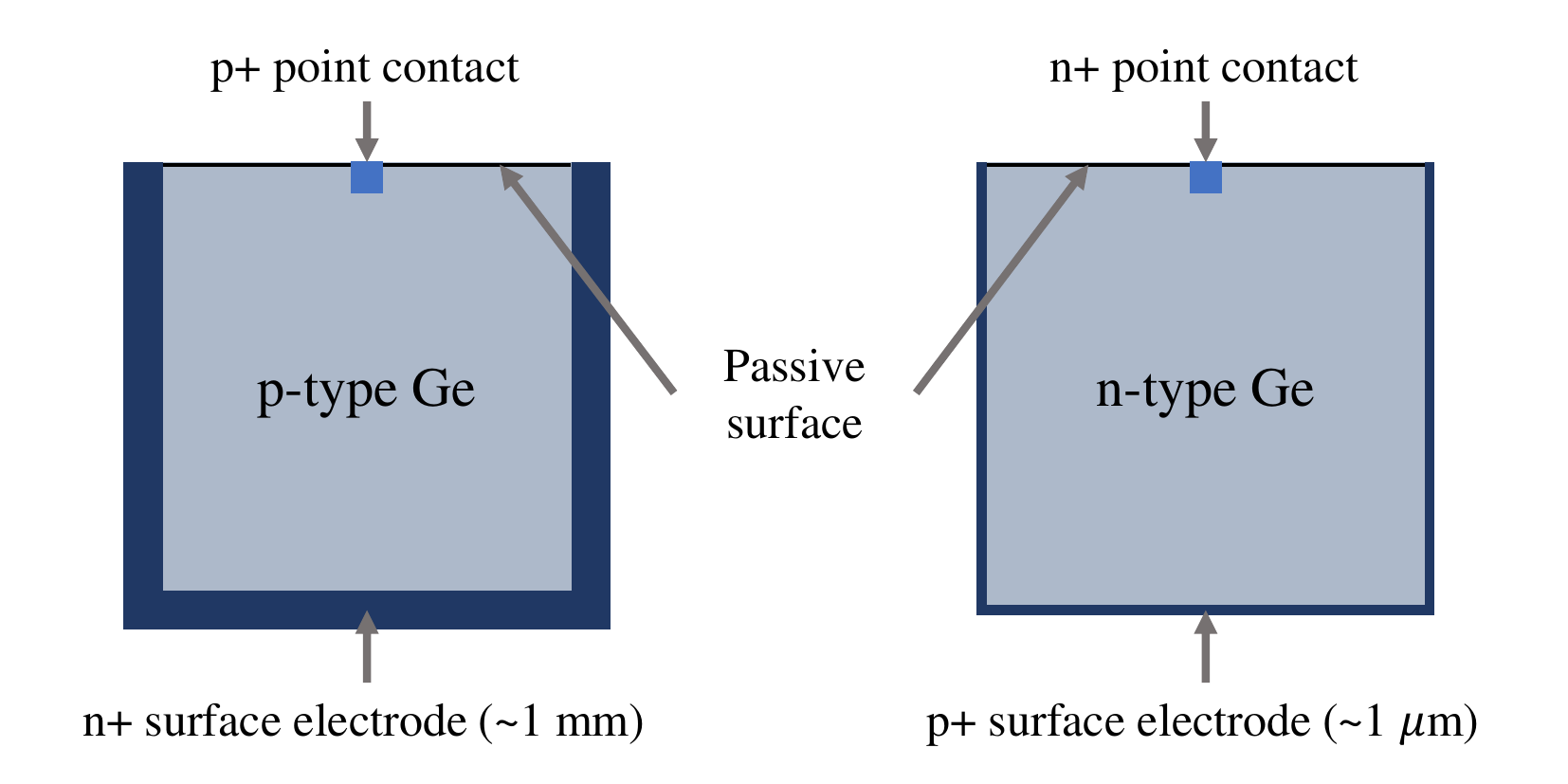}
    \caption{Schematic view of p-type~(left) and n-type~(right) PCGe detectors.}
    \label{fig:PCGe}
\end{figure}

Depending on the type of impurities within the germanium crystal, there are the original n-type PCGe~(\textit{n}PCGe) and the later realised p-type PCGe~(\textit{p}PCGe)~\cite{Barbeau2007}, which are depicted in Figure~\ref{fig:PCGe}.
As the \textit{p}PCGe exhibits non-degrading energy resolution and enhanced background rejection~\cite{Barbeau2007, Singh2019}, it is slightly more favoured amongst rare-event searches, e.g., TEXONO~\cite{Soma2016}, CDEX-1~\cite{Kang2013}, CONUS~\cite{conus2021} and CoGeNT~\cite{Aalseth2011, Aalseth2013}, etc.
Nonetheless, the relatively thicker, typically on the order of millimetres, lithium surface electrode of the \textit{p}PCGe introduces surface effects that could cause the event energy deposition to be quenched.
This leads to excess in the low-energy spectrum~\cite{Aalseth2013, LiHB2013, Wong2011}, constituting a major source of systematic uncertainty.
Incomplete account of such irregular events could potentially result in misinterpretation of the data and restrain the sensitivity of the experiment.
For instance, an anomalous excess of low-energy events was reported in CoGeNT as signatures of light weakly interacting massive particles~(WIMPs)~\cite{Aalseth2011, Aalseth2013}, contradicting the observations of TEXONO~\cite{LiHB2013, LiHB2014} and CDEX-1~\cite{Zhao2013, Yue2014}, which could very well result from an underestimate of the anomalous surface events~\cite{LiHB2014, Martin2012, Aguayo2013, Yang2018}.
It is, therefore, imperative to differentiate these surface events from regular signal events in the crystal bulk volume to correctly interpret the collected data.

\begin{figure}
    \centering
    \includegraphics[width=\columnwidth]{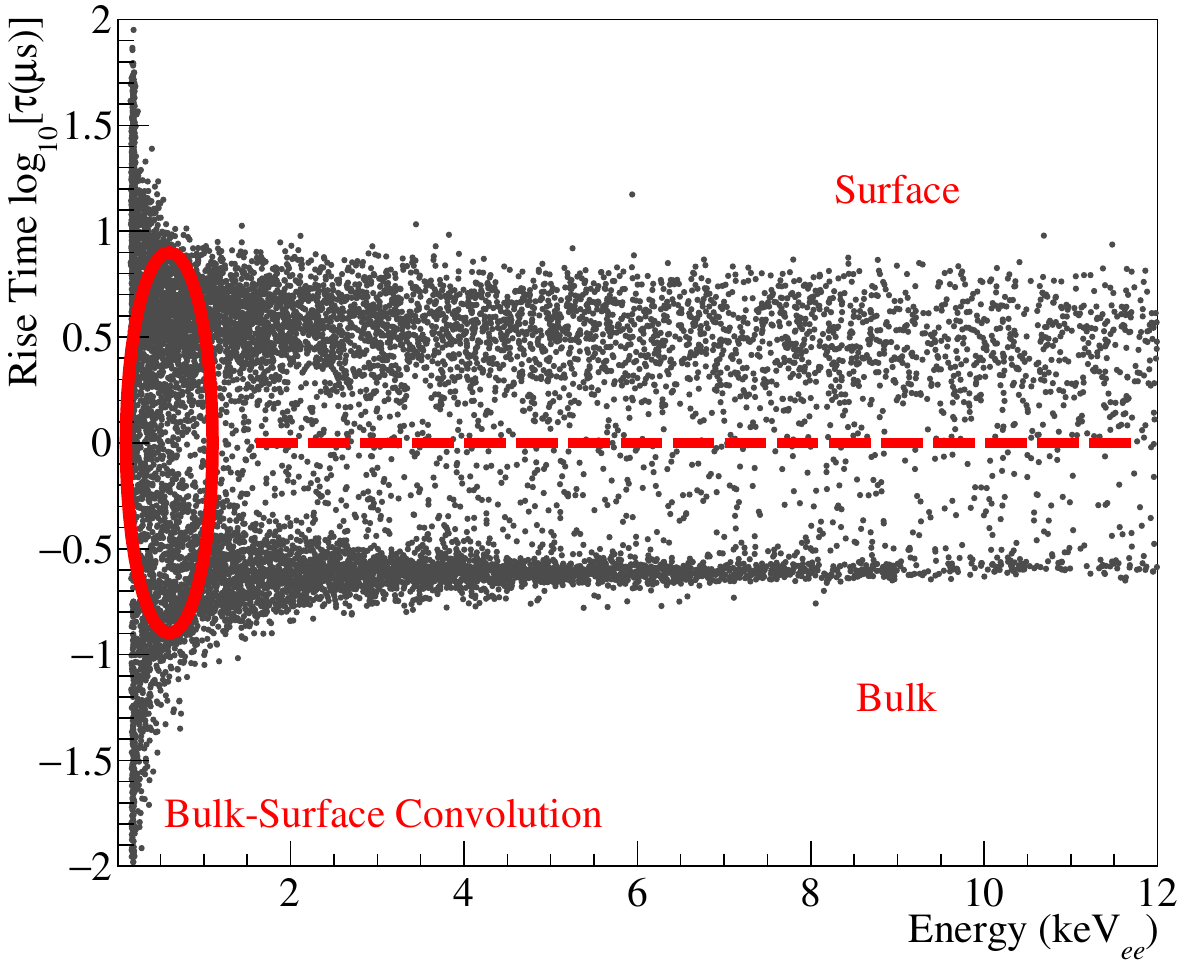}
    \caption{An example rise-time~($\mathrm{\mathrm{log_{10}}}(\tau)$) distribution plotted against energy using TEXONO data.  The plotted rise time $\tau$ is defined as the time for the pulse to rise from 5\% to 95\% amplitude. The bulk-surface convolution is indicated by the ellipse.  In fact, the two spectra are even more intertwined at low energy than one would have conjectured from this two-dimensional distribution, see Figures~\ref{fig:figure7} and \ref{fig:figure9}. }
    \label{fig:risetime2d}
\end{figure}

Surface events are usually identified by their relatively slow-rising pulse shape.
At higher energies, this approach is rather effective in differentiating the surface events from the bulk signal.
As we approach the energy threshold, however, the resolution of the pulse rise time worsens, and the rise-time distributions of the bulk and surface events starts to entangle with each other, which can be seen most clearly in Figure~\ref{fig:risetime2d}.
It is, therefore, imperative to resolve this convolution at near-threshold energies.
Earlier studies~\cite{LiHB2014, Yang2018} have provided significant insight into the problem and established the foundations of the bulk-surface discrimination technique.
In this article, we investigate in a pulse-shape fitting method based on cross-correlation shape-matching and low-pass filtering that can potentially refine the rise-time resolution and, in turn, further improve the bulk-surface discrimination technique.
The method is developed on a set of artificial pulses generated by making use of a programmable pulse generator module~(pulser).
With the truth information of these simulated pulses, we are able to evaluate the performance of the method on a more definitive basis.
Finally, we also demonstrate the effects resulting from this new method by analysing reactor data from the TEXONO experiment.

This article is organised as follows: we introduce the standard rise-time analysis together with its implications on bulk-surface discrimination in Section~\ref{sec:method}.
Section~\ref{sec:CC} then introduces the proposed pulse-shape method.
Section~\ref{sec:pulser} briefly describes the pulser setup and examines the signals generated by the pulser.
Test performances of the method on both pulser samples and TEXONO data are presented in Section~\ref{sec:perform}.
The results are summarised and discussed in Section~\ref{sec:summary}.

\section{Rise-Time Analysis}
\label{sec:method}

\begin{figure}
    \centering
    \includegraphics[width=\columnwidth]{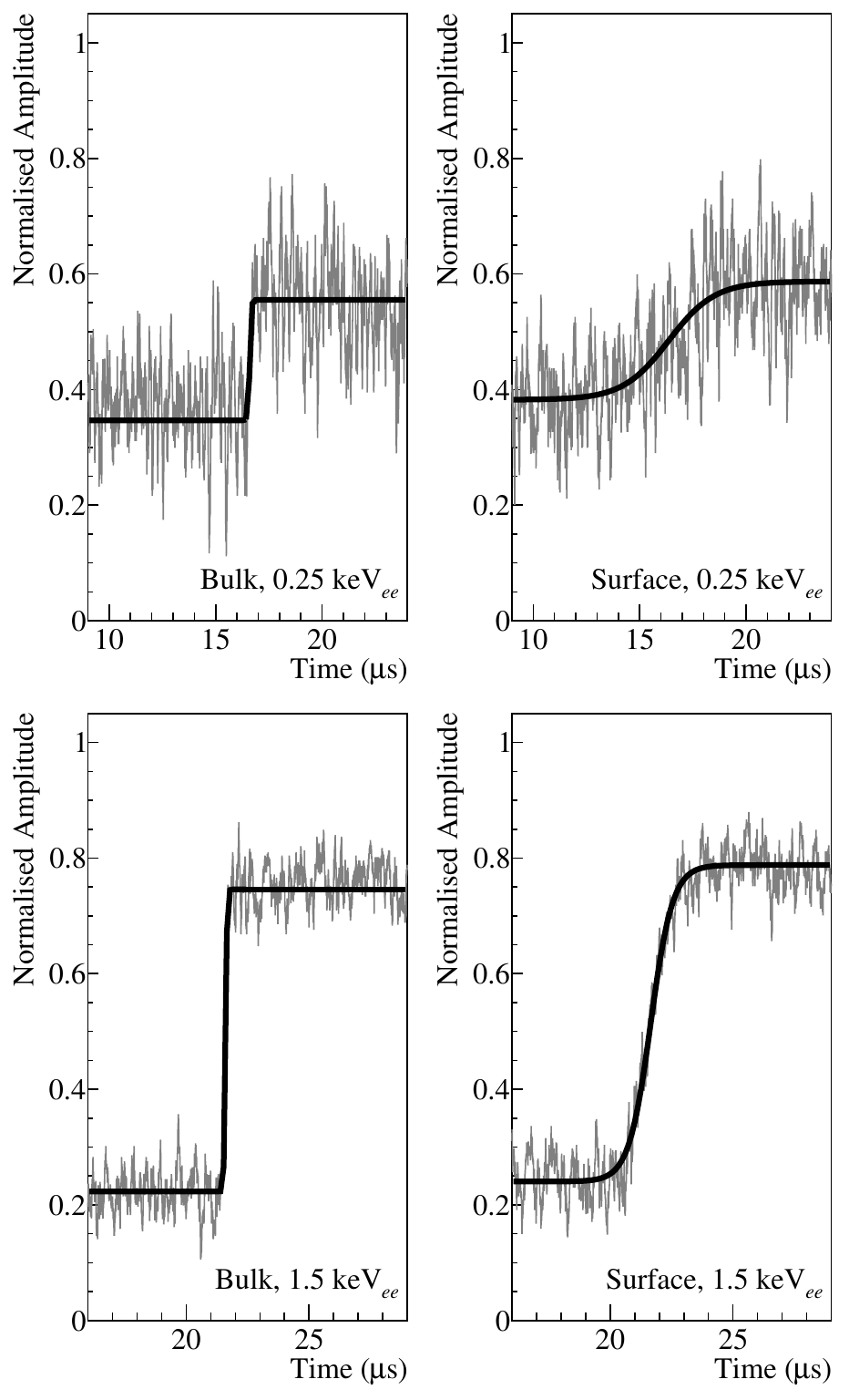}
    \caption{Typical bulk~(left) and surface~(right) pulses of energies 0.25~keV$_{ee}$~(top) and 1.5~keV$_{ee}$~(bottom) taken from the TEXONO \textit{p}PCGe detector.  Pulses are plotted alongside the fitted hyperbolic tangent curves~(black).  }
    \label{fig:raw_pulse}
\end{figure}

In PCGe detectors, the timing pulse induced by the drifting charges generally features the shape of a sigmoid curve.
To quantitatively analyse the rise time of this sigmoidal pulse, the common practice is to characterise it by a hyperbolic tangent function
\begin{equation}
  \frac{A}{2} \times \mathrm{tanh}\left( s~(t - t_0) \right) + C \, ,
\label{eq:tanh}
\end{equation}
where $A$ is the amplitude, $s$ is the time constant, $t_0$ is the time offset and $C$ is the pedestal offset. 
Figure~\ref{fig:raw_pulse} shows typical bulk and surface pulses in a \textit{p}PCGe detector at 0.25~keV$_{ee}$ and 1.5~keV$_{ee}$, where the duration of each pulse being read in is 37.5~$\mu$s.
It should be noted that the rise time $\tau$ quoted in the discussions hereafter is defined as the time for the fitted pulse shape to rise from 5\% to 95\% amplitude.
This rise time is related to the hyperbolic tangent time constant by the relation $\tau \equiv ln 19 / s$.

\subsection{Surface Effects in \textit{p}PCGe Detectors}
\label{surf:effect}
The outer surface electrode of the PCGe detector forms an inactive region, where the charge carriers experience essentially no electric field, and, hence, no charge collection.
Ideally, this constitutes a hard boundary between the inactive region and the depleted germanium bulk.
On one side, no charge can be collected, and on the other side, the charge can be fully collected.
In practice, however, atom diffusion around the boundary creates a transition region in between~\cite{Burns1990, Soma2016, Singh2019}.
The size of this transition region is negligible for the \textit{n}PCGe, owing to the thin surface layer, yet considerable for the \textit{p}PCGe as a result of the millimetre-thick lithium-diffused outer contact.
The exact properties and behaviours of this transition region has been explored in detail in previous investigations~\cite{LiHB2014} using $^{241}$Am and $^{137}$Cs $\gamma$-sources.

Charge carriers that are created within this semi-active surface are only subject to rather weak electric fields, and will, thus, have to practically diffuse through the layer to reach the depletion region, where effective charge collection is possible~\cite{Tamm1967}.
The charge diffusion time is typically on the order of several microseconds, giving rise to pulses of longer rise times.
This slower drift of the charge further provides more time for charge recombination to take place, resulting in quenching of the event energy deposition~\cite{Yang2018, Tamm1967, Ma2017}.
Therefore, the surface events generally induce weaker and slower pulses compared to that of the regular events originating from the depleted bulk region.
Furthermore, due to the loss of information through diffusion, surface events are usually considered as the background to be removed in analysis. 

\subsection{Bulk-Surface Discrimination} 

Given the slow-rising nature of the surface event pulses, the pulse rise time $\tau$ will be the main handle to differentiate between bulk and surface events.
As shown in Figure~\ref{fig:risetime2d}, the higher energy bulk and surface spectra are well separated, whereas the entanglement takes place as one enters the lower energy region.
To investigate how the spectra convolute with each other, one of the approaches~\cite{LiHB2014} was to examine the spectral shapes of the rise-time distributions for radioactive calibration sources, which are well measured.
With the calibration sources, the degree of spectral leakage arising from both the bulk and surface spectra can be derived and quantified by the background suppression efficiency and signal sacrifice.
This spectral shape study has provided pioneering insight into characterising the spectral convolution at lower energies, yet there are practical limitations to this approach.

As the relevant sacrifices and efficiencies vary along the energy spectrum, the applicability of the derived values are confined to the specific energy range covered by the calibration sources with well-known spectral shapes.
Additionally, as calibration has to be carried out periodically to ensure the stability of the analysis, the employment of external sources could lead to potential operational difficulties for long-term data collection or multi-detector experiments.
To address these limitations of the spectral shape method, the ratio method was further developed and proposed in~\cite{Yang2018}. 
The ratio method exploits the fact that the ratio between the rates of different event sources are, in principle, solely dependent on energy and fixed with respect to the rise-time spectrum.
This allows for the use of in situ data, rather than external radioactive sources of known spectral shapes, for calibration, and, hence, significantly expands the scope of applicability of the bulk-surface discrimination analysis.
It has also been confirmed that the resolution provided by the ratio method is consistent with that of the spectral shape method.

These approaches currently provide the most comprehensive account for bulk-surface spectral leakage in the low-energy region at the rise-time distribution level.
The key to further enhance the bulk-surface discrimination power will, thus, rely on the refinement of the pulse rise-time resolution.
This requires advancements at the pulse-fitting level to more accurately determine the rise time $\tau$ itself and fundamentally reduce the degree of spectral leakage, which is the focus of this study.

\section{Proposed Pulse-Shape Method}
\label{sec:CC}

The hyperbolic tangent fit involves a total of four parameters, as in Equation~\ref{eq:tanh}.
A common feature of multi-parameter fits is that the results are usually quite sensitive to the initial parameter setting, i.e., seeding.
Since the time constant $s$, and hence the derived rise time $\tau$, is the result to be delivered by the hyperbolic tangent fit, we will limit our focus to the initial evaluation of the other three parameters in the fit, namely, the time offset $t_0$, the amplitude $A$ and the pedestal offset $C$.
From past experience, it was noticed that amongst the parameters, the seed of the time offset $t_0$ is of most influence, and can very well feed back to the initial estimation of the other parameters.

The existing procedure adopted by the TEXONO and CDEX-1 experiments, proposed in~\cite{LiHB2014}, for obtaining the seeding for the parameters is rather straightforward.
It starts out by smoothing the raw signal pulse with the Savitzky-Golay filter~\cite{Golay1964}.
This smoothed pulse shape is then fitted with the hyperbolic tangent function, from which the fit results are taken as the seed for the actual fits on the raw pulse.
This method will, thus, be designated as the smoothed fit method hereafter.

For higher energies, the smoothed fit method generally provides satisfactory estimates, yet there is still room for improvement as one approaches lower energies.
In this section, we propose an algorithmic approach that utilises cross-correlation matching combined with low-pass filtering to enhance the accuracy of initial parameter estimates.

\subsection{Cross-Correlation Matching}

Another approach to determining the time offset $t_0$ of the pulse is via shape-matching.
Based on some a priori assumptions or information regarding the signal pulses, it is, in principle, possible to construct a reference pulse shape that is to be matched with the signal.
At the point of best match, usually represented by the successful extremisation of some similarity metric, the known information of the reference shape can be said to represent that of the signal pulse.
In the specific case of the PCGe timing pulses, one would be matching the signal pulses with a hyperbolic tangent-like reference shape, which can be, for instance, a hyperbolic tangent of some given characteristic time constant or the averaged pulse shape, etc.
The time offset estimate would then be the $t_0$ of the reference shape at best match.
Cross-correlation is one such metric that has been widely adopted in the astrophysical community for studies of stellar~\cite{optics1985, optics2002, Stellar2017} and galactic~\cite{Galaxy2006} optical observations, pulsar timing~\cite{pulsar1998}, gravitational waves~\cite{GW1999, LIGO2004} and supernovae~\cite{SN2020}, etc.
The main innovation in our adaption of this conventional similarity metric to the pulse-shape studies for PCGes lies in the construction of the reference shape, which is vital for facilitating the practical application of the method.

\begin{figure}
    \centering
    \includegraphics[width=\columnwidth]{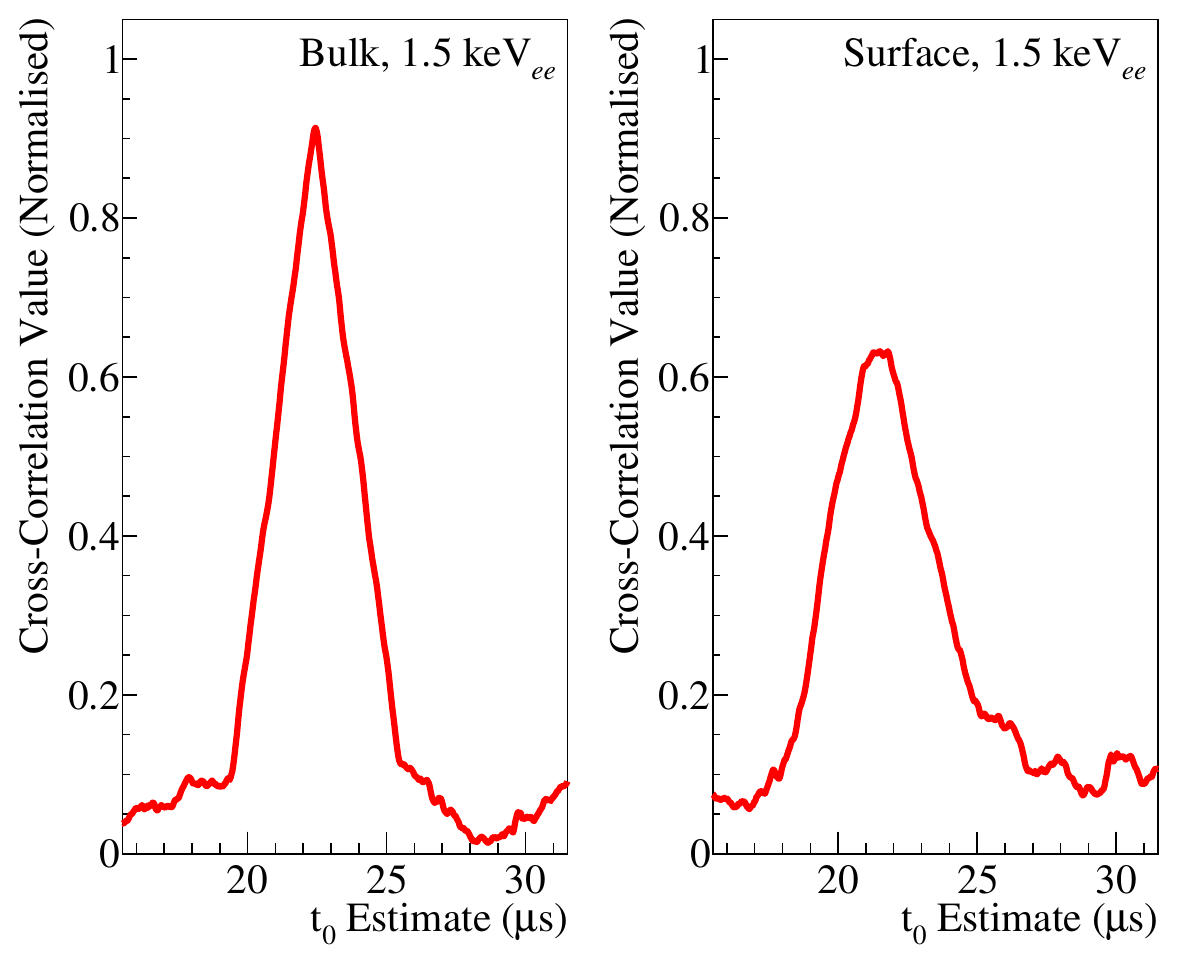}
    \caption{The cross-correlation profile for bulk~(left) and surface~(right) events of 1.5~keV$_{ee}$ energy using a step reference pulse. }
    \label{fig:cc_profile}
\end{figure}

The cross-correlation between the signal pulse and reference shape takes the form of
\begin{equation}
  \mathcal{C}(\Delta t) \equiv \int_{t_i +\Delta t -\frac{T}{2}}^{t_i +\Delta t + \frac{T}{2}} R(t-\Delta t) \cdot S(t) \mathrm{d}t \, ,
\label{eq:cc}
\end{equation}
where $\mathcal{C}$ stands for the cross-correlation value, $\Delta t$ is the time displacement of the reference shape with respect to some initial position in time $t_i$, $T$ is the width of the integration window, $R$ stands for the reference shape and $S$ stands for the signal pulse.
Here the reference shape is scanned through a range of different $t_0$ values represented by $t_i + \Delta t$, considering the case when the reference shape is held at the centre of the integration window.
At each step of this process, the integral will be carried out as in Equation~\ref{eq:cc}, eventually forming a profile as in Figure~\ref{fig:cc_profile}.
The $t_0$ estimate is then given by the position that corresponds to the maximum cross-correlation value.
It should be noted that, in practice, the signal pulses are comprised of discrete points, and, hence, the cross-correlation values are evaluated as a discrete sum
\begin{equation}
  \mathcal{C}(\Delta t) = \frac{1}{N_T} \sum_{j = 0}^{N_T-1} R_j \cdot S_{j+\Delta t} \, ,
\end{equation}
where $\Delta t$ is the time displacement taking on only integer values, $N_T$ is the total number of points included within the summation window, $R_j$ and $S_j$ are now the discrete series of the reference shape and signal pulse, respectively.
The $t_0$ value corresponding to each $\Delta t$ is then $N_T/2 + \Delta t$ in the discrete case.

By examining the integral in Equation~\ref{eq:cc}, it is not difficult to conclude that the cross-correlation of two hyperbolic tangent functions maximises when the $t_0$ of the two functions coincide.
Thus, justifying the arbitrariness of the choice of reference shape amongst hyperbolic tangent-like shapes.
Nonetheless, we shall argue that a step function rising from $-1$ to $1$, viz., a hyperbolic tangent of infinite time constant and zero pedestal offset, is the optimal choice for the reference shape.
As it can be observed from Figure~\ref{fig:cc_profile}, the cross-correlation profile peak is rounded and widened for longer rise times.
This can also be confirmed analytically by carrying out the integral in Equation~\ref{eq:cc}.
In fact, the analytical result suggests that the resolution of the peak is mainly determined by the longer rise time of the two integrands.
The step function is, therefore, an appealing choice on theoretical grounds as it does not have a finite rise time to convolute with that of the signal pulse.
Another advantage, on the more practical end, of this choice is that it is independent of the signal pulse, allowing a more automated and streamlined procedure for the analysis to be possible.
It is also worth noting that the pedestal offset of the reference shape affects the resulting cross-correlation profile, so it is necessary to keep it fixed at zero.
The amplitude, on the other hand, has no effect on the result, and can, therefore, take on any positive value, i.e., it can be any step function rising from $-n$ to $n$ where $n \in \mathbb{R}^+_\ast$.

\subsection{Low-Pass Filter}

\begin{figure}
    \centering
    \includegraphics[width=\columnwidth]{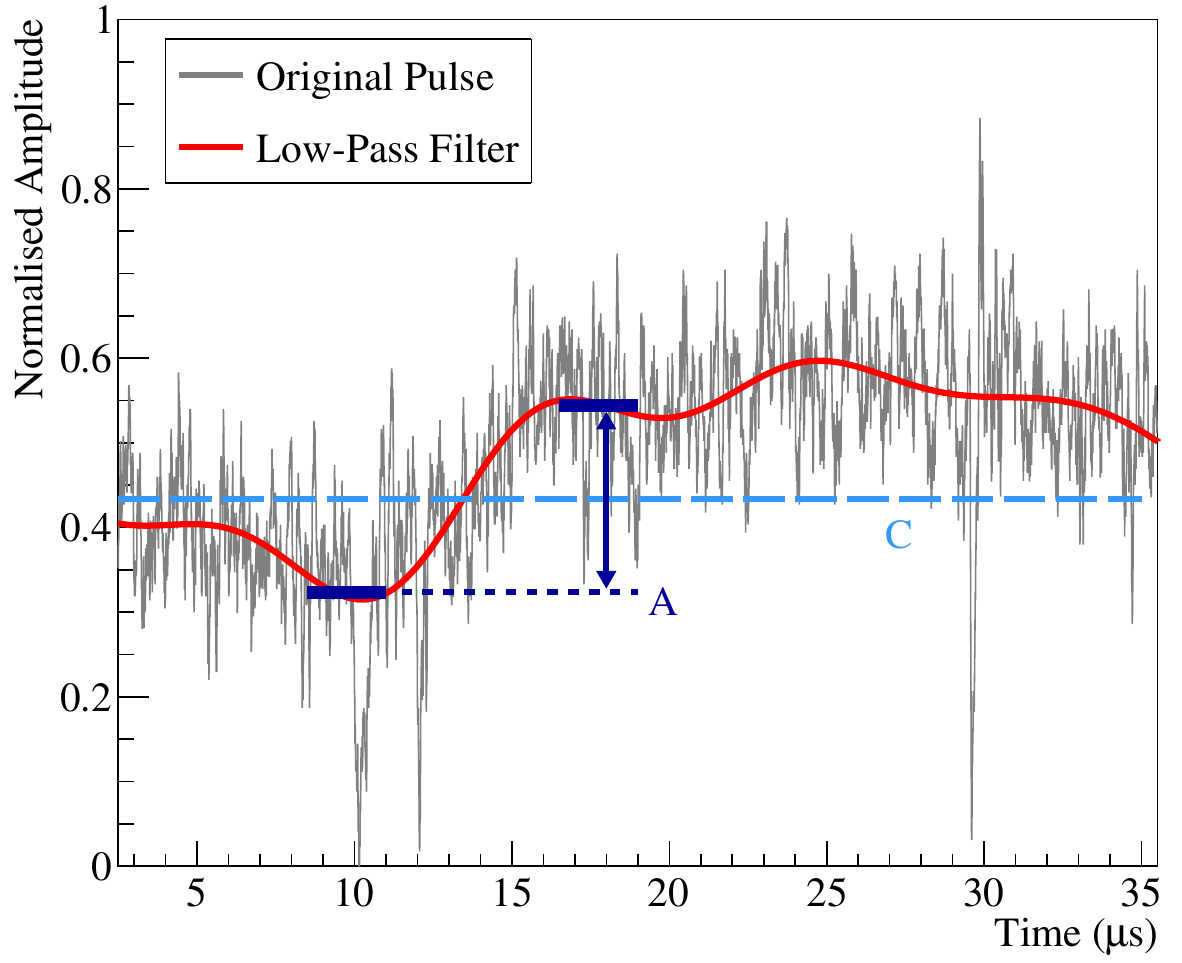}
    \caption{An example of a noisy pulse~(grey) at 0.25~keV$_{ee}$ plotted against the output of the low-pass filtered result~(red), where the time segments chosen for evaluation of the limiting values~(navy) are also plotted together with the corresponding $C$~(dashed blue) and $A$~(navy arrow) estimates.   It is evident that the low-pass filter sketches out the underlying trend and suppresses the aberrant behaviours due to the electronic noise.  }
    \label{fig:figure5}
\end{figure}

With the time offset $t_0$ estimate determined, the remaining parameters are the amplitude $A$ and pedestal offset $C$.
Given the sigmoidal behaviour of the pulse, the evaluation of the amplitude and pedestal offset will rely on the determination of the asymptotic values at both ends, with $A$ being their difference and $C$ being their mean, as illustrated in Figure~\ref{fig:figure5}.
This can intuitively be done by taking the average of the end segments of the pulse, e.g., a common practice is to calculate the mean value of the pulse from 2.5 to 5~$\mu$s and from 32.5 to 35~$\mu$s.
Nonetheless, the fluctuations caused by the electronic noise often leads to noticeable deviations in the evaluation of the asymptotic values.
It will, therefore, be advantageous to apply smoothing techniques when performing this estimate.

By inspecting the general behaviour of the pulses, it is transparent that the sigmoidal behaviour is dictated by the low-frequency components, whereas the noise fluctuations are generally of higher frequency.
A natural choice would then be to filter out the high-frequency components, i.e., applying a low-pass filter to the pulse.
This is, of course, at the risk of losing fast-varying information, for instance, rise-time evaluations on the filtered pulse will result in overestimates as the fast components are suppressed.
Nevertheless, since the objective is confined to the determination of limiting values of the pulse, this should not constitute a problem.
Figure~\ref{fig:figure5} illustrates the effect of the low-pass filter, where one can see how it reveals the underlying information encompassed in the pulse.

Furthermore, as the pulse in Figure~\ref{fig:figure5} suggests, the amplitude estimate from a simple average at the ends can deviate from the true value, even with the low-pass filter applied, due to the trends in the noise structure.
In general, as one departs further from $t_0$, the pulse levels are more likely to be distorted by the interim noise structure.
To minimise this effect, one should identify the time segments which are reasonably close to $t_0$ yet has already passed the sloping region, e.g., the 2.5~$\mu$s after reaching 95\% amplitude and before rising to 5\% amplitude.
This designation of time segments, however, varies with the rise time $\tau$ of the pulse.
Since the rise time of each pulse cannot be known a priori, a conservative approach is to assume a rise time below which covers 99\% of the surface events. 
In this study, this upper bound for rise time $\tau$ is chosen to be 6~$\mu$s, and, hence, the corresponding time segments will be 3 to 5.5~$\mu$s before and after $t_0$, as in Figure~\ref{fig:figure5}, where the estimate for $t_0$ will be obtained from the cross-correlation calculation.
It should also be noted that since the improvement in the $t_0$ estimate is the most influential, this proposed algorithm will, henceforth, be referred to as the cross-correlation method for convenience.

\section{Pulser-Generated Signals}
\label{sec:pulser}

\begin{figure}
    \centering
    \includegraphics[width=\columnwidth]{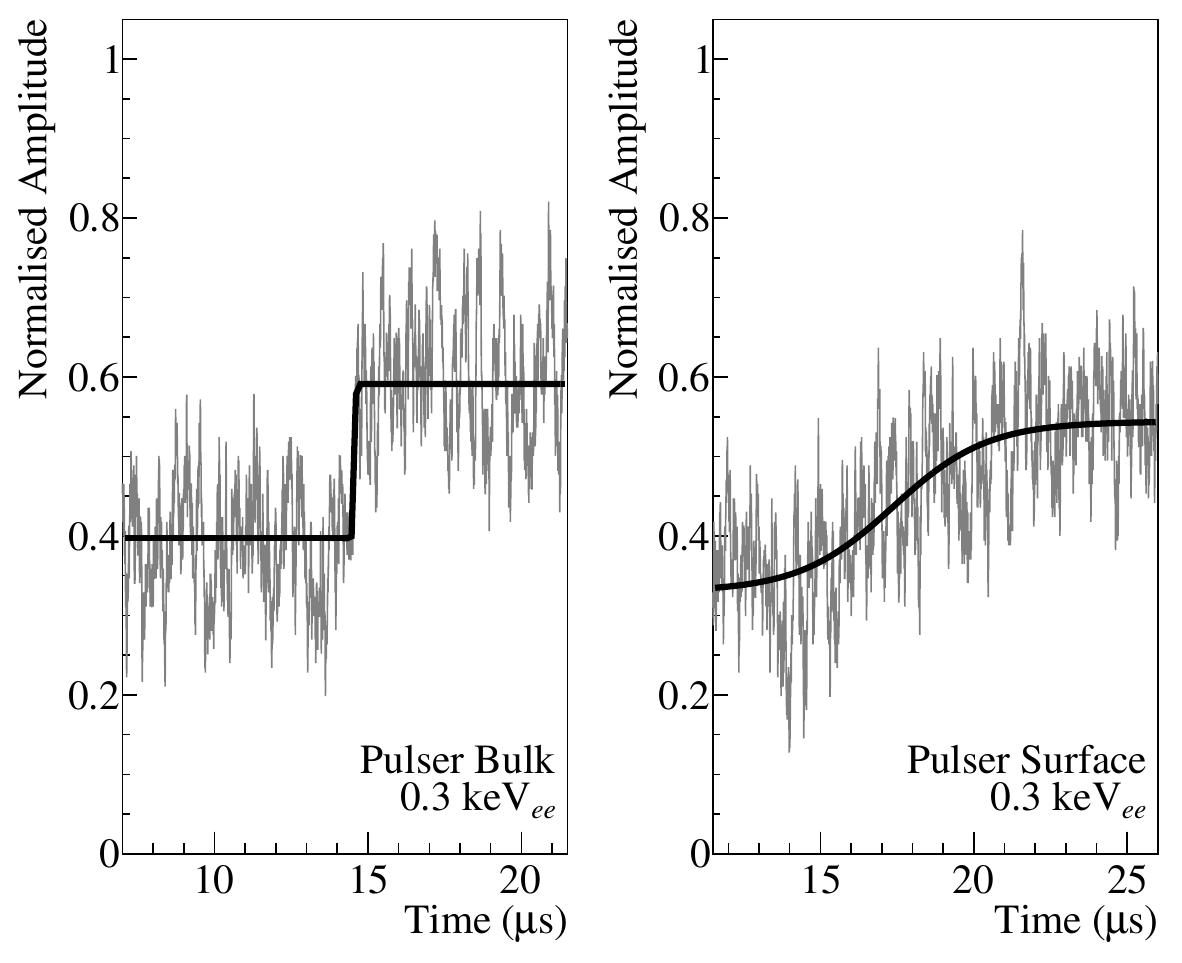}
    \caption{Pulser-generated bulk~(left) and surface~(right) pulses of energy 0.3~keV$_{ee}$ collected from the \textit{n}PCGe detector at Academia Sinica.  Pulses are plotted alongside the fitted hyperbolic tangent curves~(black). }
    \label{fig:figure6}
\end{figure}

\begin{figure}
    \centering
    \includegraphics[width=\columnwidth]{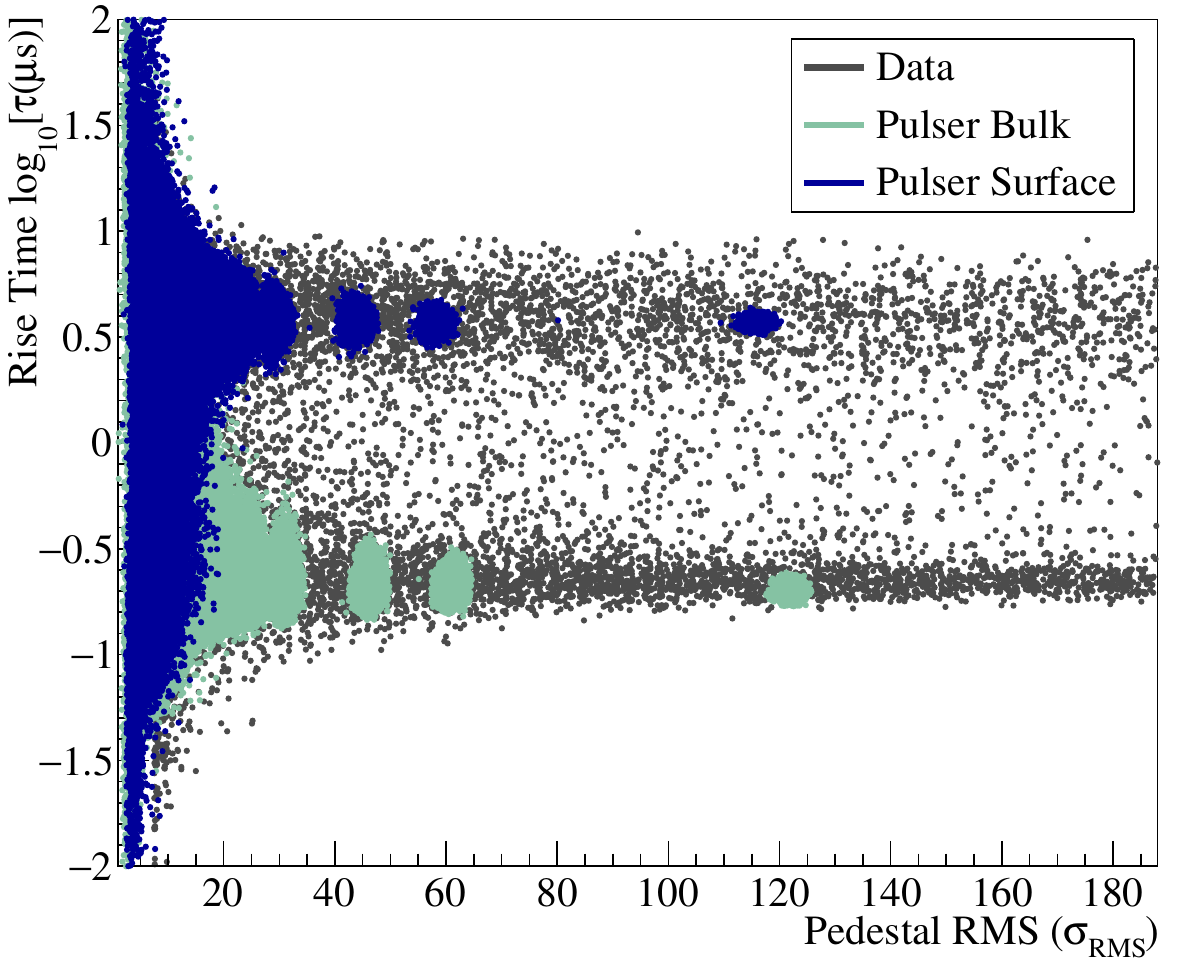}
    \caption{The rise-time~($\mathrm{\mathrm{log_{10}}}(\tau)$) distribution plotted against energy, comparing TEXONO data~(grey) and pulser-generated bulk~(green) and surface~(navy) samples from Academia Sinica.  The plotted rise time $\tau$ is defined as the time for the pulse to rise from 5\% to 95\% amplitude.  The TEXONO data is collected with \textit{p}PCGe and pulser samples with \textit{n}PCGe.   As a result, the energy is expressed in units of pedestal-noise-profile RMS~($\sigma_{\mathrm{RMS}}$), which is the root mean square obtained from the electronic-noise profile subject to each detector.  The analysis threshold generally corresponds to 7~$\sigma_{\mathrm{RMS}}$. }
    \label{fig:figure7}
\end{figure}

Separate samples of bulk and surface events are necessary for the correct evaluation of the effectiveness of pulse-shape methods.
Though calibration sources, as in~\cite{LiHB2014}, can provide comprehensive information regarding the rise-time distributions of the bulk and surface spectra at certain energy ranges, a homogeneous sample covering the whole energy spectrum will have to rely on artificial signals.
It is, in principle, possible to generate such samples via software approaches by simulating the particle behaviours in the entire $p$PCGe configuration.
However, the current accuracies of this ab initio approach generally does not match that of the calibration-based or data-driven approaches, as discussed in~\cite{Yang2018}.
Alternatively, as in~\cite{Conus2024}, one can produce artificial pulses imitating different signal pulses by using a programmable pulse generator module~(pulser).
The pulser can generate pulses of designated rise time at arbitrary energies that will be directly fed to the pre-amplifier for further data acquisition.
A more realistic pulse sample incorporating the detector electronic noise can then be generated.
Figure~\ref{fig:figure6} shows examples of bulk-like and surface-like pulses generated by the pulser at 0.3~keV$_{ee}$.

For the purpose of taking pulser-generated samples, we employ the pulser module PXI-5412, produced by National Instruments, connected to an \textit{n}PCGe detector at Academia Sinica.
Since the lower energy rise-time spectrum of bulk and surface events are yet to be resolved, the pulser-generated pulse settings are tuned to resemble the higher energy bulk and surface spectra in data.
With such settings, the pulser-generated pulses should be representative of the behaviour of bulk and surface events at lower energies when the input voltage is decreased.
Figure~\ref{fig:figure7} shows the rise-time distribution plotted against energy for pulser-generated pulses compared to real data.
Note that since the data and pulser samples are taken by \textit{p}PCGe and \textit{n}PCGe detectors, respectively, the energy in Figure~\ref{fig:figure7} is expressed in the pedestal noise level as a universal basis for comparison between different detectors.
Also, one might notice that the pulser-generated surface-like spectrum is much narrower than in data.
This is due to the simple fact that a single time constant is used in the pulser setting, whereas surface events in data could take on a range of different rise times depending on the site of interaction within the transition layer.
It is rather transparent from Figure~\ref{fig:figure7} that the pulser sample agrees with data quite well at higher energies, and the low-energy behaviour of the pulser sample also seems to be overall in line with data.
This justifies our use of the pulser-generated bulk-like and surface-like pulse samples as the test sample for pulse-shape methods.

\section{Performance}
\label{sec:perform}

With the pure pulse samples generated by the pulser, we are in the position to examine the performance of our proposed pulse-shape method.
In this section, we will present, first of all, the improvements in estimating the pulse time offset $t_0$ by using the cross-correlation method compared to the existing smoothed fit method.
The influence on the overall rise-time analysis yielded by this improvement in seeding will then be demonstrated with the pulser samples.
Finally, the method is tested on real data from the TEXONO experiment as a final assessment of effectiveness.

\begin{figure}
    \centering
    \includegraphics[width=\columnwidth]{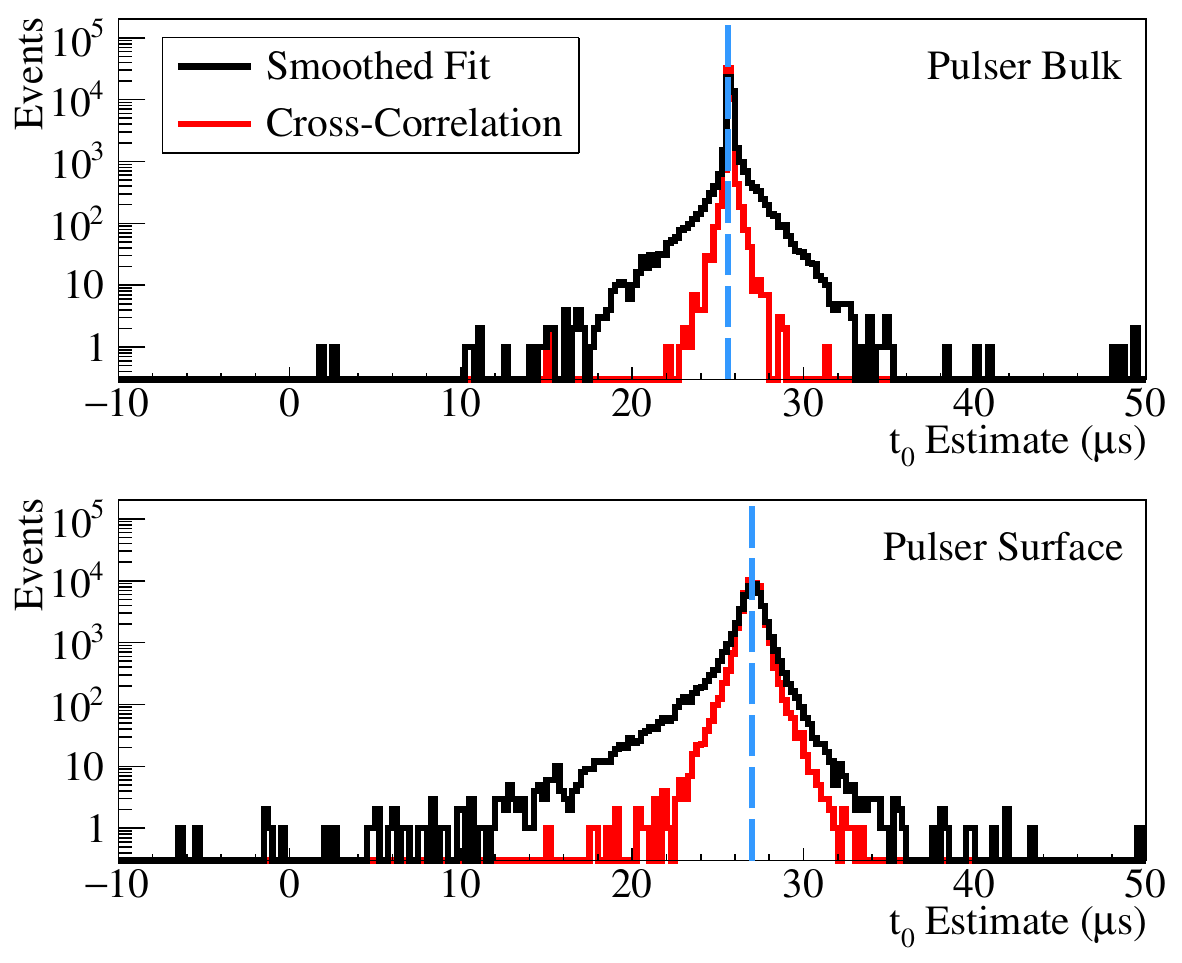}
    \caption{$t_0$ estimate distributions of the pulser-generated bulk~(top) and surface~(bottom) events obtained by the cross-correlation~(red) and smoothed fit methods~(black).  The distribution covers the energy range from 0.3~keV$_{ee}$~(near detection threshold) to 0.8~keV$_{ee}$, above which the two methods provide similar results.  The dashed blue line indicates the true $t_0$ programmed in the pulser. }
    \label{fig:figure8}
\end{figure}

\subsection{Performance in \texorpdfstring{$t_0$}{TEXT} Estimate}

Figure~\ref{fig:figure8} shows the $t_0$ estimate distribution of the pulser-generated bulk-like and surface-like samples obtained by the cross-correlation method compared to that of the smoothed fit method.
Events from the near-threshold 0.3~keV$_{ee}$ to 0.8~keV$_{ee}$, above which the two methods are in good agreement, are included in the distribution.
With the true time offset shared amongst pulser-generated pulses being known, we are able to conclude that the precision in determining $t_0$ is substantially refined by applying the cross-correlation method at lower energies.

\subsection{Effect on Rise-Time Analysis}
\label{sec:CC_risetime}

\begin{figure}
    \centering
    \includegraphics[width=\columnwidth]{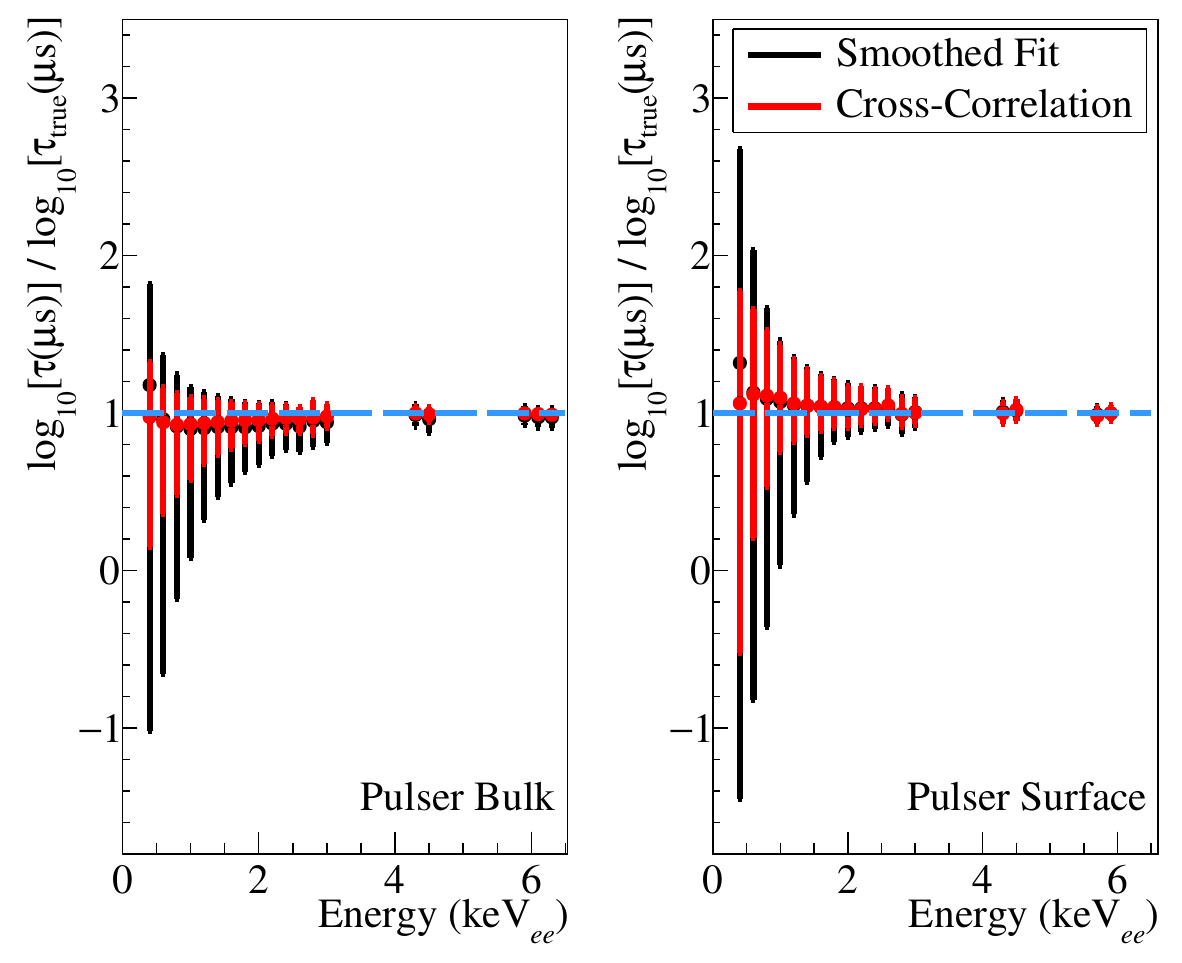}
    \caption{The normalised rise-time~($\mathrm{log_{10}}(\tau) \, / \, \mathrm{log_{10}(\tau_{\mathrm{true}})}$) mean and uncertainty plotted against energy for pulser-generated bulk~(left) and surface~(right) events using the cross-correlation~(red) and smoothed fit~(black) methods.  The normalisation constant $\mathrm{log_{10}}(\tau_{\mathrm{true}})$ is the logarithm of the true rise time $\tau_{\mathrm{true}}$ that was programmed in the pulser, and, hence, unity~(dashed blue) indicates the absence of bias.  The error bars extending to the negative range, on the other hand, indicates the degree of spectral leakage into the other spectrum, i.e., surface leakage into the bulk spectrum and vice versa. }
    \label{fig:tau_bias}
\end{figure}

\begin{figure}
    \centering
    \includegraphics[width=\columnwidth]{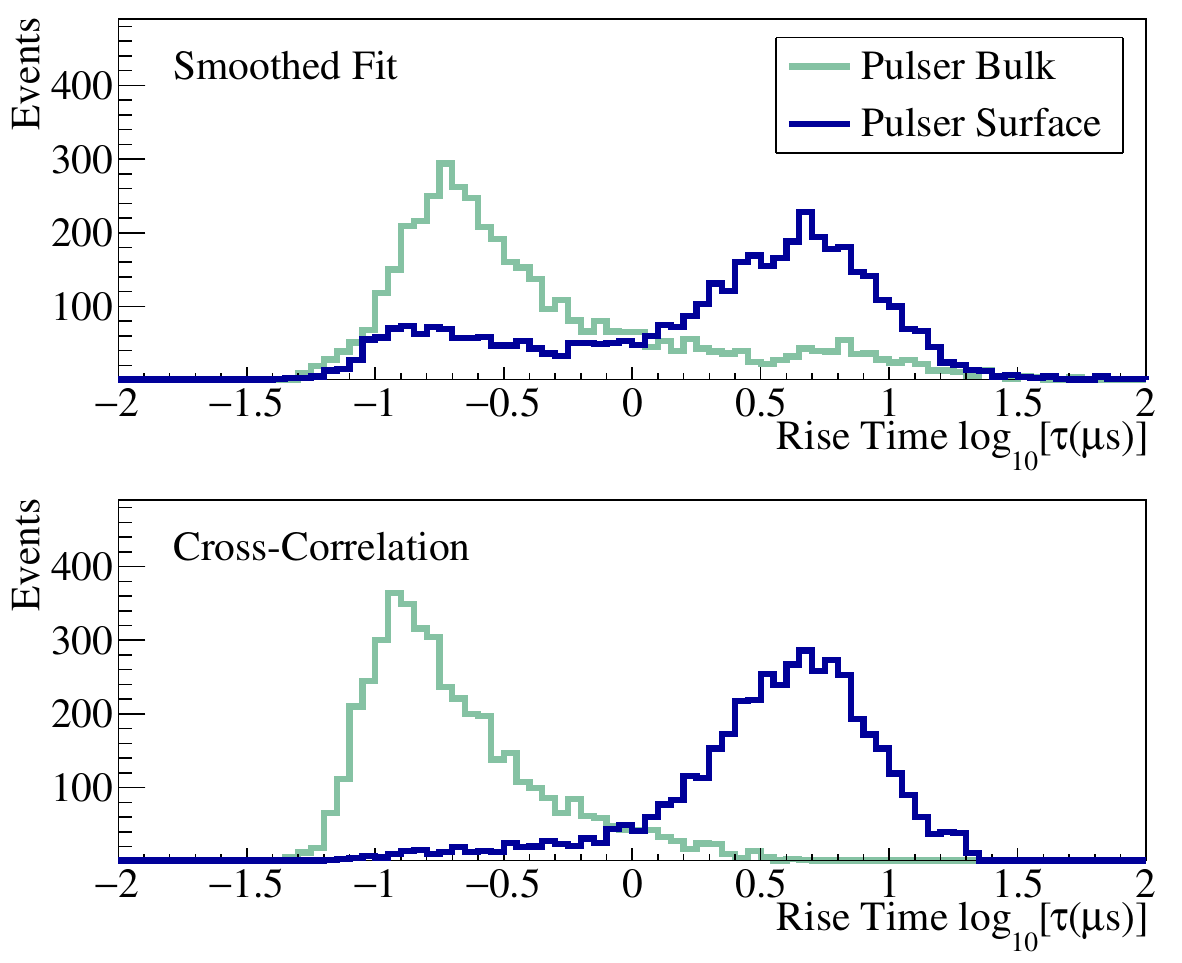}
    \caption{The rise-time~($\mathrm{log_{10}}(\tau)$) distribution of the pulser-generated bulk~(green) and surface~(navy) events using the cross-correlation~(top) and smoothed fit~(bottom) methods at 0.3~keV$_{\mathrm{ee}}$~(roughly 7~$\sigma_{\mathrm{RMS}}$, near detection threshold).  The most noticeable improvement brought by the cross-correlation method is the elimination of the false-fit peaks.  }
    \label{fig:figure9}
\end{figure}

\begin{figure}
    \centering
    \includegraphics[width=\columnwidth]{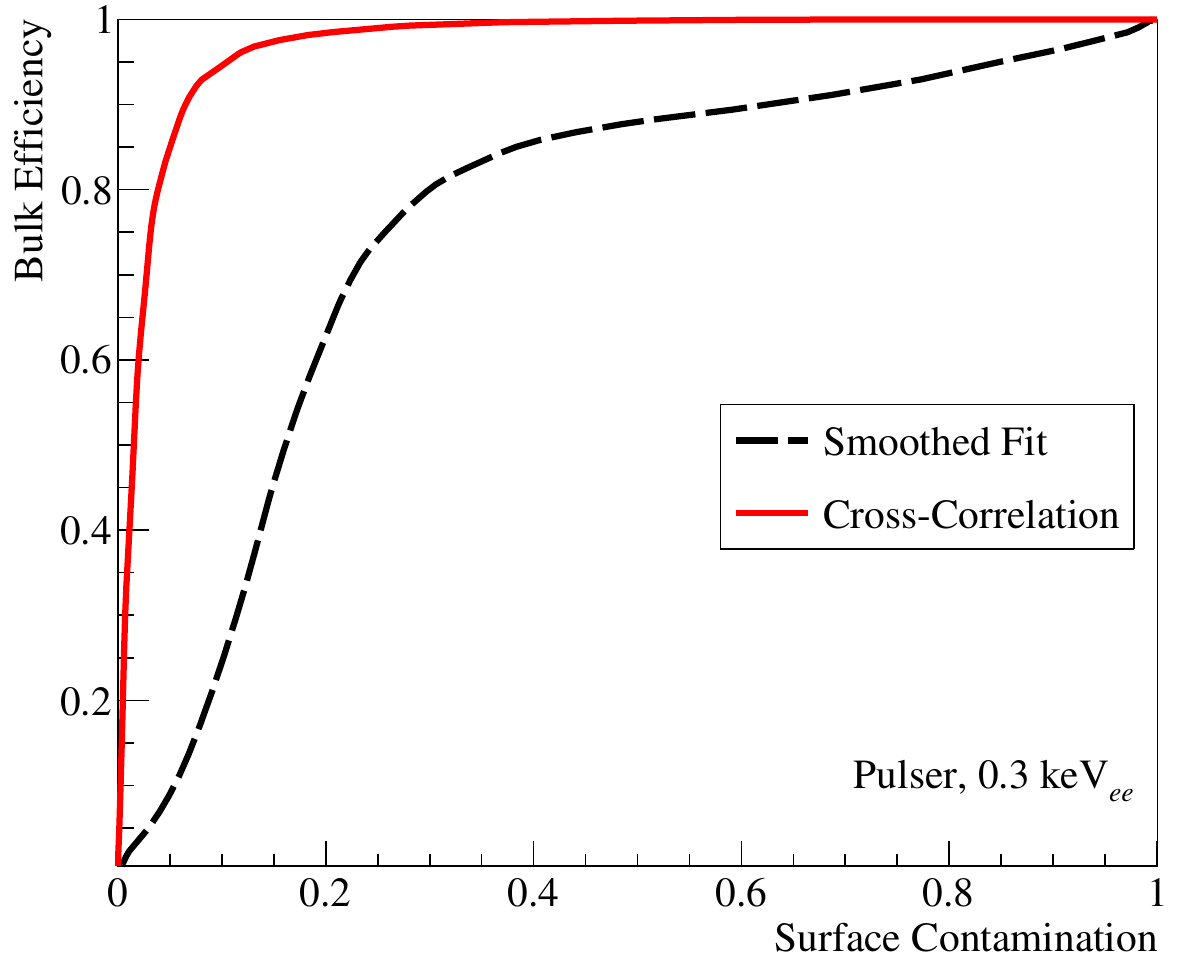}
    \caption{The receiver-operation-characteristic~(ROC) curve of the cross-correlation~(red) and smoothed fit methods~(dashed black) for bulk-surface discrimination at 0.3~keV$_{ee}$~(roughly 7~$\sigma_{\mathrm{RMS}}$, near detection threshold).  The $x$-axis shows the fraction of surface events identified as bulk events by a given rise-time cut, whereas the $y$-axis shows the fraction of bulk events correctly identified by that given cut. }
    \label{fig:roc_curve}
\end{figure}

The cross-correlation method is then applied to the rise-time analysis for obtaining the initial seeding for the fit parameters, employing, again, the same pulser-generated test samples.
Figure~\ref{fig:tau_bias} shows the resulting normalised rise time~($\mathrm{log_{10}}(\tau) \, / \, \mathrm{log_{10}(\tau_{\mathrm{true}})}$) mean and uncertainty plotted against energy, where the logarithm of the rise time $\tau$ is normalised by which of the programmed true rise time $\tau_{\mathrm{true}}$ of the pulser. 
It is immediately evident that overall the biases are not significant, confirming the fidelity of both fitting methods, and the main effect of the new approach is reflected in the substantially reduced uncertainty at lower energies.
To further visualise this improvement, the rise-time distributions for events at 0.3~keV$_{ee}$, near the detection threshold, are shown in Figure~\ref{fig:figure9}.
The most noticeable effect of the improved $t_0$ estimation is the elimination of the false-fit peak, i.e., the secondary peak caused by the surface event leakage into the rise-time spectrum of the bulk peak and vice versa.
The size of the false peak of both the surface and bulk events are reduced by more than 70\%.
Furthermore, the primary peaks are generally more concentrated than in the previous version of the fit, and the overall distributions are also more constrained.
This apparent improvement can be visualised even better by plotting the receiver-operation-characteristic~(ROC) curve as in Figure~\ref{fig:roc_curve}.
In the ROC curve, each point corresponds to a specific cut value on rise time for bulk event selection, where its $y$ coordinate indicates the fraction of bulk events preserved by this cut and its $x$ coordinate records the fraction of surface events leaking in.
For example, the cut that identifies bulk events by $\mathrm{log_{10}}(\tau) < 0$ would correspond to the point~(0.09, 0.95) on the cross-correlation ROC curve.
Figure~\ref{fig:roc_curve}, thus, clearly demonstrates the overall enhancement in bulk-surface discrimination power with the new approach.

Another advantage worth noting is that alongside the improvement in rise-time resolution, the number of rise-time fit iterations, in order to deliver reasonable resolution, can also be reduced.
In the existing procedure, the rise-time fit is iterated three times for each pulse, not including the smoothed fit.
This can be time-consuming from the viewpoint of data processing.
With the new approach, it appears that the gain from performing the third iteration is marginal.
Additionally, since the time spent on the cross-correlation calculation is almost negligible compared to that of a fit, the number of fits can be brought down from, practically, 4~(one smoothed fit plus three raw-pulse fits) to 2.
Thus, nearly halving the computing time of the rise-time analysis.

\subsection{Test on TEXONO Data}
\label{sec:texono_data}

Following the signs of improvements observed in the preliminary investigations on the pulser-generated test samples, the cross-correlation method is then implemented into the TEXONO data analysis to confirm the effectiveness and robustness of the proposed algorithm.
The TEXONO experiment~\cite{TEXONO2018}, using a 1.434~kg $p$PCGe detector with fiducial mass 1.383$\pm$0.014~kg, is located at the Kuo-Sheng Reactor Neutrino Laboratory~(KSNL) with a 30~m.w.e. overburden.
In addition, the $p$PCGe detector is enclosed by the NaI(Tl) crystal scintillators as anti-Compton~(AC) suppressor and plastic scintillator panels for cosmic-ray~(CR) veto.
For TEXONO data analyses, the events of interest are, therefore, the ones that do not trigger the AC or CR shielding, denoted AC$^{-}\otimes$CR$^{-}$, potentially arising from sources like neutrinos or dark matter as well as cosmogenic background or self-triggered electronic noise.
Since the aim of the discussion in this section is to verify the functionality of the algorithm and to investigate the potential impact on the analysis threshold, currently near 0.2~keV$_{ee}$, our attention will mainly be confined to the AC$^{-}\otimes$CR$^{-}$ events.

\begin{figure}
    \centering
    \includegraphics[width=\columnwidth]{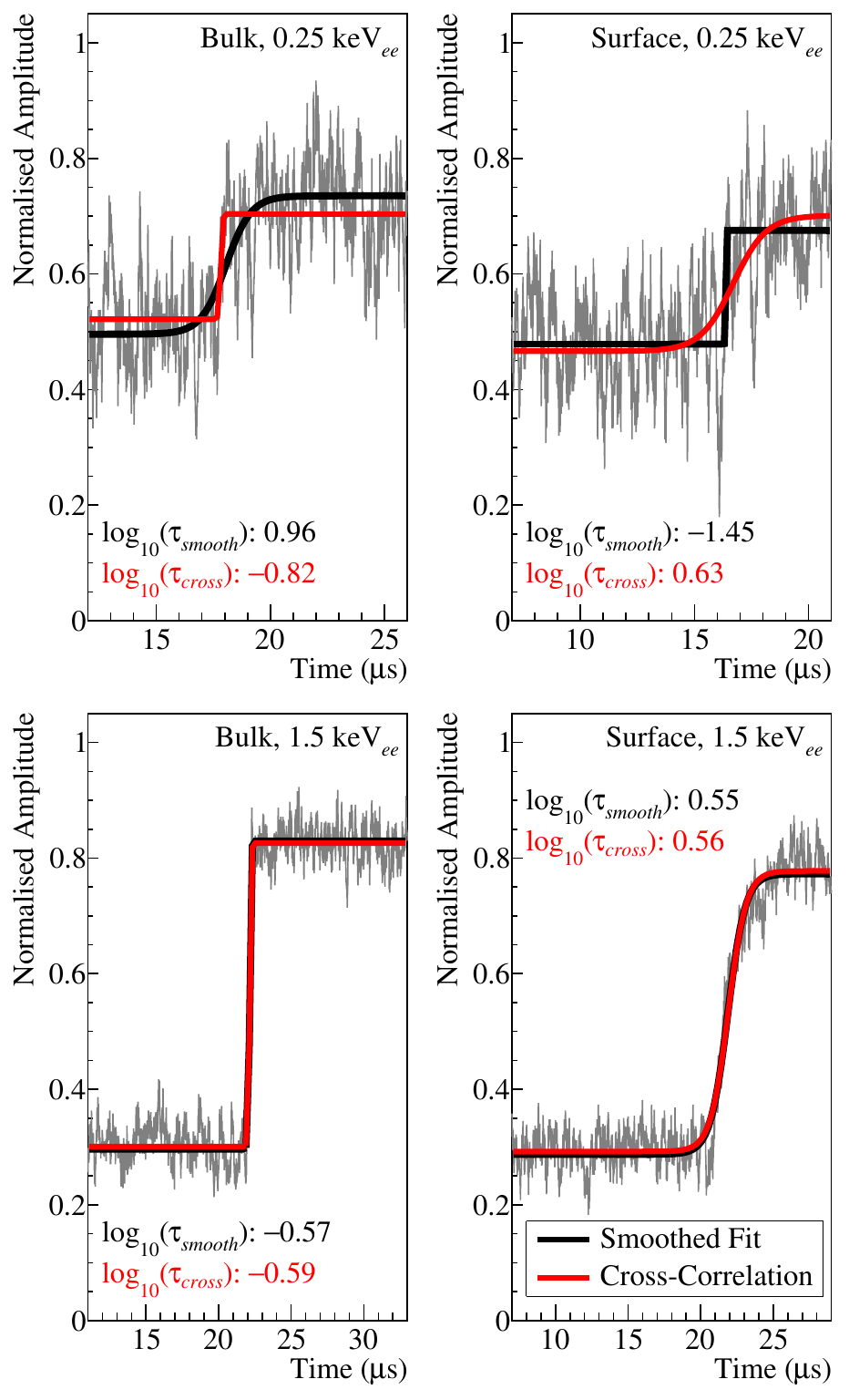}
    \caption{Typical bulk~(left) and surface~(right) pulses of energies 0.25 keV$_{ee}$~(top) and 1.5 keV$_{ee}$~(bottom) taken from the TEXONO $p$PCGe detector.  Pulses are further plotted alongside the fitted hyperbolic tangent curves using the smoothed fit~(black) and cross-correlation~(red) methods.  It is evident that for lower energies, the cross-correlation method is able to provide a more reasonable estimate of the pulse rise time.  }
    \label{fig:figure11}
 \end{figure}

\subsubsection{Refined Rise-Time Resolution}

Figure~\ref{fig:figure11} shows the fitting results of typical data pulses at 0.25~keV$_{ee}$ and 1.5~keV$_{ee}$.
It is clear that the cross-correlation method produces results that are congruous with the smoothed fit method in the higher energy region, reassuring the reliability of the new algorithm.
On the other hand, with the implementation of the cross-correlation method, an immediate improvement in the rise-time evaluation of lower energy events can also be observed.
The new algorithm appears to be able to determine the pulse rise time more accurately at the pulse level, and, hence, better discriminate between bulk and surface events.
Examples of the cross-correlation method correcting the rise-time estimation of misidentified low-energy pulses are exhibited in the top panels of Figure~\ref{fig:figure11}.

\begin{figure}
    \centering
    \includegraphics[width=\columnwidth]{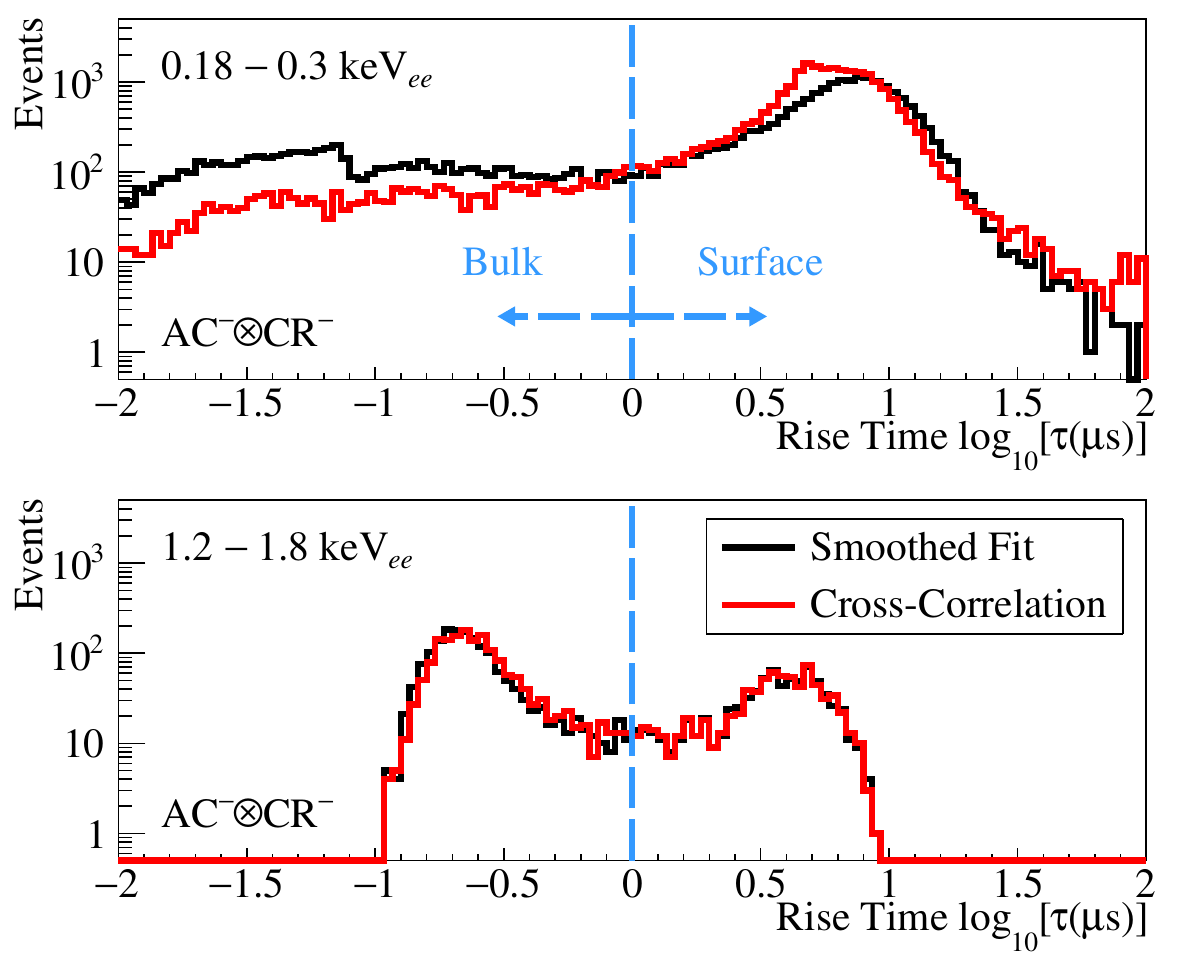}
    \caption{The rise-time~($\mathrm{log_{10}}(\tau)$) distribution of AC$^{-}\otimes$CR$^{-}$ events at energies from 0.18~keV$_{ee}$ to 0.3~keV$_{ee}$~(top) and 1.2~keV$_{ee}$ to 1.8~keV$_{ee}$~(bottom) derived from the smoothed fit~(black) and cross-correlation~(red) methods using 49~kg-days of TEXONO reactor data collected at KSNL.  The dashed blue line indicates the boundary between bulk and surface events.  It is evident that though the methods agree well at higher energies, the cross-correlation method significantly reduces the low-energy surface event leakage into the bulk region.}
    \label{fig:figure12}
\end{figure}

The effect of this more appropriate identification of bulk and surface events is even more evident at the rise-time distribution level.
Figure~\ref{fig:figure12} shows the rise-time distributions from a 49~kg-days~(roughly equivalent to 34~live days) TEXONO dataset at energies from 0.18~keV$_{ee}$ to 0.3~keV$_{ee}$ and 1.2~keV$_{ee}$ to 1.8~keV$_{ee}$ using the smoothed fit and cross-correlation methods.
As usual, the two methods agree fairly well at higher energies.
At near-threshold energies, however, the results differ rather significantly.
It is observed that the number of events in the bulk spectrum is reduced by nearly 70\% with the cross-correlation method compared to the smoothed fit approach.
Though ostensibly quite different, this behaviour is, in fact, in accordance with the earlier observations of false-fit peak suppression on the pulser-generated test samples shown in Figure~\ref{fig:figure9}.
In the pulser discussion of section~\ref{sec:CC_risetime}, the sample sizes of bulk and surface events are roughly the same, whereas in data the excess of surface events dominates the lower energy spectrum due to the quenched energy deposition, which is consistent with earlier observations in~\cite{Aalseth2013, LiHB2013, Wong2011}.
As a result, the false-fit peak resulting from the leakage of surface events, as in Figure~\ref{fig:figure9}, will overwhelm the low-energy bulk spectrum, creating a threshold below which the signal spectrum will be completely smeared.
The removal of the surface leakage from the bulk spectrum is, therefore, a sign of improvement in the determination of pulse rise times, and leads to the possibility of probing towards lower energies in analysis.

\begin{figure}
    \centering
    \includegraphics[width=\columnwidth]{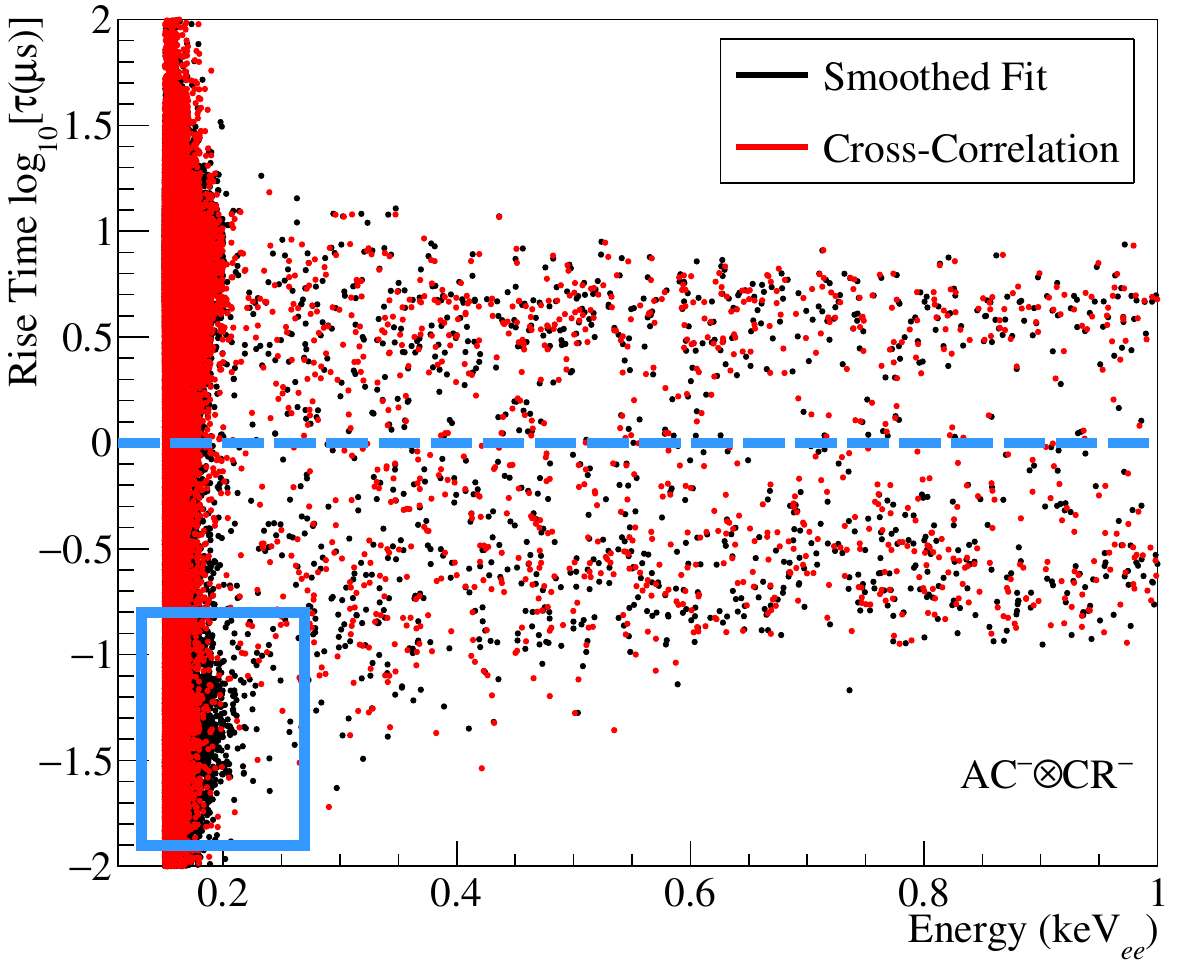}
    \caption{The rise-time~($\mathrm{log_{10}}(\tau)$) distribution of AC$^{-}\otimes$CR$^{-}$ events using the smoothed fit~(black) and cross-correlation~(red) methods plotted against energy.  The dense region in the bulk spectrum roughly corresponds to the analysis threshold.  Therefore, the box region cleared out by the cross-correlation method indicates the possibility of lowering the analysis threshold.  }
    \label{fig:figure13}
\end{figure}

This improvement can be further visualised in the rise-time distributions plotted against energy, as shown in Figure~\ref{fig:figure13}.
Note that Figure~\ref{fig:figure13} displays a smaller data sample, as opposed to Figure~\ref{fig:figure12}, such that the density scales are visually discernible.
From this distribution, the choice of the analysis threshold of 0.2~keV$_{ee}$ becomes transparent as it corresponds to the boundary near the noise edge, at which the number of microphonic noise events rise steeply.
The effect of the cross-correlation approach then shows its strength by eliminating much of the contamination at the near-threshold bulk region, causing the rising edge of events to recede and, thus, potentially lowering the analysis threshold.

\subsubsection{Impact on Analysis Threshold}
\label{en:spec}

\begin{figure}
    \centering
    \includegraphics[width=\columnwidth]{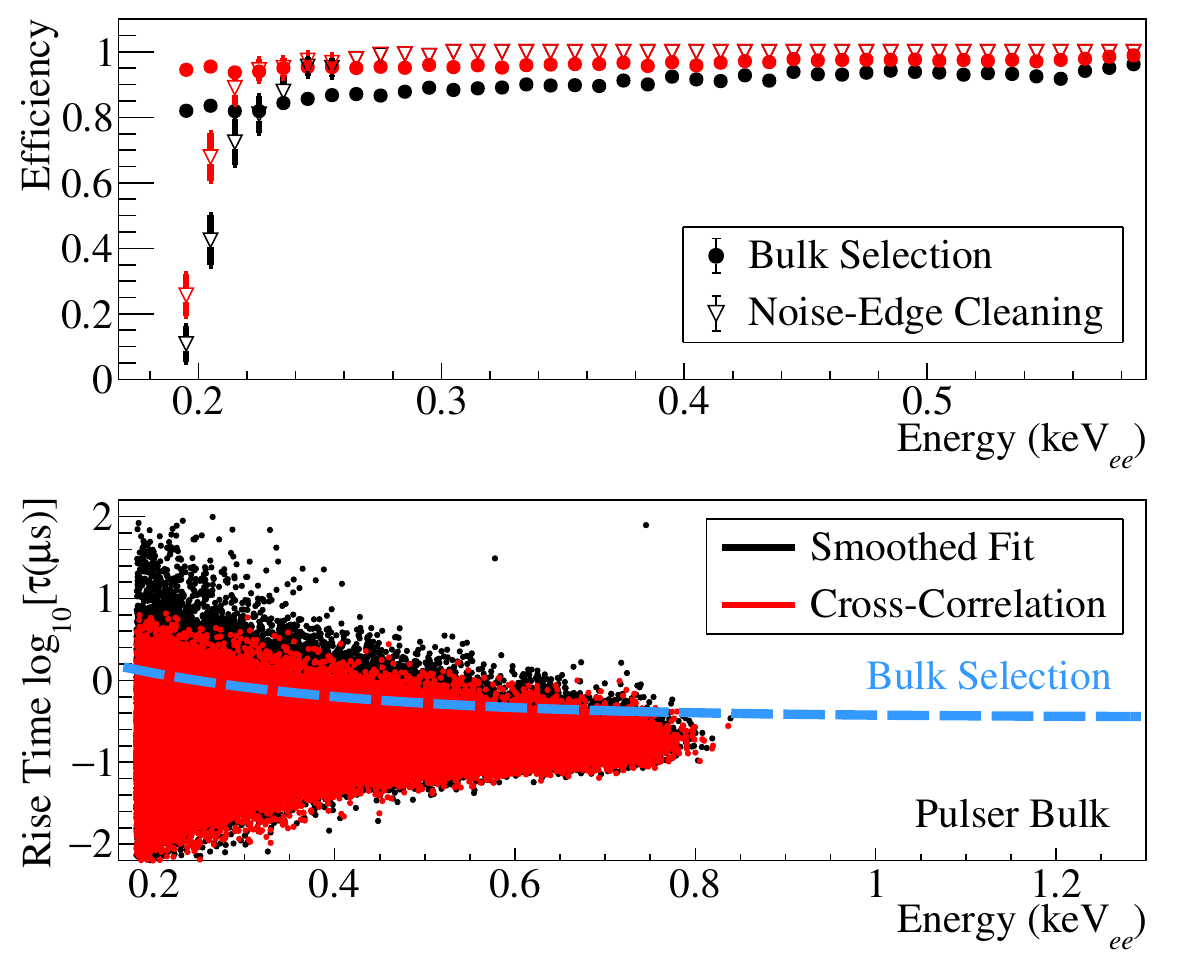}
    \caption{The rise-time~($\mathrm{log_{10}}(\tau)$) distribution of pulser-generated bulk sample using the smoothed fit~(black) and cross-correlation~(red) methods plotted against energy~(bottom).  The dashed blue line indicates the bulk signal selection curve, below which the events are regarded as measured bulk events B$_{\mathrm{m}}$.  Alongside the rise-time versus energy spectrum, the corresponding efficiencies of the bulk selection~(solid circle) and signal retention of noise-edge cleaning cuts~(hallow inverted triangle) are also shown~(top).  }
    \label{fig:figure14}
\end{figure}

The discussion on the analysis threshold is hitherto mainly based on qualitative observations.
In order to establish a rigorous quantitative assessment for the impact of the cross-correlation method on the analysis threshold, a full analysis on the energy spectrum is necessary.
Compared to the basic treatments in the rise-time analysis, the full energy spectrum analysis takes on a few additional treatments in signal selection and noise reduction.

First of all, instead of assigning events as bulk and surface according to the rather simplistic reference of $\mathrm{log_{10}}(\tau) = 0$, the pulser can, again, be employed to study the bulk spectral shape for the optimisation of signal selection.
The bottom panel of Figure~\ref{fig:figure14} shows the bulk-like rise-time distribution generated by deploying the pulser to the KSNL $p$PCGe detector, where the dashed blue curve represents the optimised energy-dependent bulk selection.
Events below the bulk selection curve are designated as the measured bulk events~(B$_{\mathrm{m}}$).
The efficiency of this selection is then plotted out against energy in the top panel of Figure~\ref{fig:figure14}, from which it can be observed that the selection efficiency of the cross-correlation method is consistently above 95\% and greater than that of the smoothed fit approach over the entire energy range of interest.
This result, thus, re-emphasises the more refined resolution in rise-time evaluation of the cross-correlation algorithm.

Furthermore, the excess in self-triggered electronic noise events rises significantly as one approaches the noise edge, which eventually renders the spectrum unanalysable.
Cleaning cuts are, therefore, implemented to reduce the level of the microphonic noise and identify the actual noise edge in analysis~\cite{Singh2019, CDEX2018}.
The cleaning cuts are designed by exploiting the correlations between the various amplifiers in the TEXONO data acquisition~(DAQ) system, including the Canberra-2026 shaping amplifiers of 6~$\mu s$ and 12~$\mu s$ shaping times, for energy measurements, and the Canberra-2111 timing amplifier, for rise-time measurements.
For physics events, the amplitudes and measured energies of the pulses derived from each amplifier generally follow a linear relation.
The self-triggered electronic noise, on the contrary, are usually less correlated and exhibits rather eccentric behaviours in these parameter spaces, which can, thus, be used to suppress the microphonic noise level.
Alongside the bulk selection efficiency, the top panel of Figure~\ref{fig:figure14} also shows the signal retention efficiency associated with the noise-edge cleaning cuts plotted against energy, where the retention efficiency is increased as a result of the more fidelitous evaluation of the timing-pulse amplitude.

\begin{figure}
    \centering
    \includegraphics[width=\columnwidth]{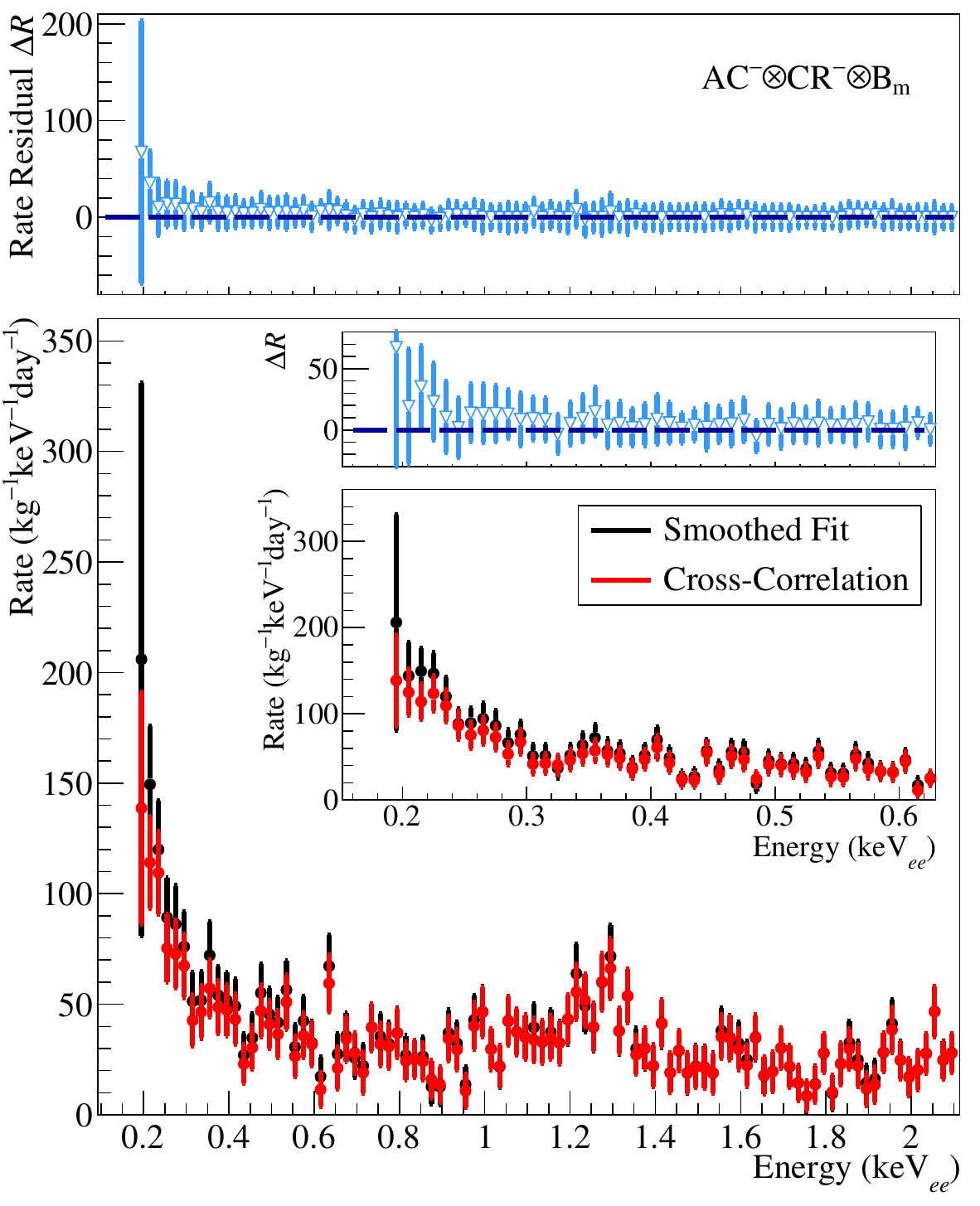}
    \caption{The measured energy spectrum~(main bottom) of AC$^{-}\otimes$CR$^{-}\otimes$B$_{\mathrm{m}}$ events derived from the smoothed fit~(black) and cross-correlation~(red) methods and the associated rate residual spectrum~(main top), where the dashed navy line indicates the null value.  Note that the inset panels show the spectra within a finer energy range.  A set of 49~kg-days TEXONO reactor data collected at KSNL is used.  The rate residuals are defined as $\Delta R \equiv R_{\mathrm{smooth}} - R_{\mathrm{cross}}$, where $R_{\mathrm{smooth}}$ and $R_{\mathrm{cross}}$ stands for the rates derived from the smoothed fit and cross-correlation methods, respectively.  It is clear from the spectra that the cross-correlation method shows strength in reducing the background level and, thus, effectively lowering the analysis threshold.}
    \label{fig:figure15}
\end{figure}

\begin{table}
  \centering
  \caption{A comparison of the first six bins in the energy spectra, shown in Figure~\ref{fig:figure15} (more clearly in the inset panels), derived from the smoothed fit and cross-correlation methods. The unit ``cpkkd'' stands for ``counts per kg-keV-day'', providing a normalised measure for the event rate with respect of the detector mass, energy range and exposure time.  The efficiencies of the noise-edge cleaning cut and bulk selection are denoted as $\varepsilon_{\mathrm{NE}}$ and $\varepsilon_{\mathrm{B_{m}}}$, respectively.  }
  \setlength{\tabcolsep}{0.7em} 
\begin{center}
\renewcommand{\arraystretch}{1.3}
\begin{tabular}{ccccc}
\hline \hline
{\bf Energy}         & Rate                              & Uncertainty                         & $\varepsilon_{\rm{NE}}$ & $\varepsilon_{\rm{B}_{m}}$  \\ 
(keV$_{ee}$)    & (cpkkd)                           & ($\pm$ cpkkd)                       & (\%)                  & (\%)            \\ \hline
\multicolumn{5}{c}{{\bf Smoothed Fit}}   \\ \hline 
{\bf 0.19}          & 206                               & 125                                 & 11 $\pm$5.5           & 82$\pm$0.8      \\  
{\bf 0.20}          & 144                                & 38                                  & 42 $\pm$7.7           & 83$\pm$0.7      \\ 
{\bf 0.21}          & 149                                & 26                                  & 72 $\pm$7.0           & 82$\pm$0.8      \\  
{\bf 0.22}          & 147                                & 24                                  & 80$\pm$5.6           & 82$\pm$0.8      \\ 
{\bf 0.23}          & 120                                & 22                                  & 87$\pm$4.5           & 84$\pm$0.8      \\ 
{\bf 0.24}          & 88                                & 18                                  & 95$\pm$2.8           & 85$\pm$0.7      \\ \hline 
\multicolumn{5}{c}{{\bf Cross-Correlation}}   \\ \hline
{\bf 0.19}          & 139                               & 53                                  & 26$\pm$6.0            & 95$\pm$0.5      \\  
{\bf 0.20}          & 125                                & 27                                  & 68$\pm$7.2            & 95$\pm$0.4      \\ 
{\bf 0.21}          & 114                                & 21                                  & 89$\pm$4.9            & 96$\pm$0.4      \\  
{\bf 0.22}          & 124                                & 20                                  & 95$\pm$3.2           & 95$\pm$0.4      \\ 
{\bf 0.23}          & 110                                & 19                                  & 96$\pm$3.0           & 95$\pm$0.4      \\ 
{\bf 0.24}          & 87                                & 16                                  & 98$\pm$2.2           & 96$\pm$0.3      \\ \hline \hline

\end{tabular}
\end{center}
\label{tab::num}
\end{table}

Finally, with the bulk selection and noise reduction cut optimised, the energy spectrum of AC$^{-}\otimes$CR$^{-}\otimes$B$_{\mathrm{m}}$  events using, again, the 49~kg-days TEXONO reactor dataset collected at KSNL can then be calculated, as shown in Figure~\ref{fig:figure15}.
Alongside the rates, the difference between the methods are further demonstrated with the rate residual~($\Delta R$), defined as the signal rate excess in the smoothed fit approach compared to the cross-correlation approach.
The residual spectrum is depicted in the top panel of Figure~\ref{fig:figure15}.
From the residual spectrum, it is apparent that the residual value starts to significantly deviate from null, i.e., the new algorithm effectively suppresses the surface leakage, for energies below approximately 0.4~keV$_{ee}$, consistent with the observations thus far.
For more detailed comparisons between the smoothed fit and cross-correlation approaches at lower energies, the measured rates alongside the relevant efficiencies of the first six bins in the energy spectra are listed in Table~\ref{tab::num}.
By inspecting the numbers listed in Table~\ref{tab::num}, apart from the obvious suppression of surface event leakage and reduced uncertainties, a considerable increase on the order of 10\% can also be observed in both the signal retention efficiency and the bulk selection efficiencies.
It then becomes immediately evident that with the enhanced performance in rise-time resolution, the lowest energy bin in the table is now rendered analysable.
This can roughly be translated to the lowering of the analysis threshold by at least 10~eV$_{ee}$, from the original 200~eV$_{ee}$ to 190~eV$_{ee}$.

\section{Summary and Prospects}
\label{sec:summary}

We have presented the performance of the proposed new pulse-shape discrimination method based on cross-correlation calculations combined with a low-pass filter for the study of the timing amplifier pulse rise times in point-contact germanium~(PCGe) detectors.
By applying the cross-correlation method on pulser-generated samples and reactor data from the TEXONO experiment, we have demonstrated that this new approach offers an overall enhancement to our abilities in conducting the rise-time analysis.
Furthermore, the consistent improvements across different detectors, including the TEXONO $p$PCGe detector at KSNL and the local $n$PCGe test detector at Academia Sinica, confirms the robustness of the algorithm and reflects its potentiality for wider applications within the field of germanium experiments.
Here we will briefly summarise the primary effects on data analysis at different levels associated with the implementation of the proposed method:

\begin{enumerate}
\item Improved initial parameter estimation:
At the lowest level of the rise-time analysis, the fidelity of the hyperbolic tangent fit is largely determined by the initial seeding of the parameters $t_0$, $A$ and $C$.
The results in Figures~\ref{fig:figure5} and \ref{fig:figure8} have clearly illustrated the substantial enhancements in  $t_0$ determination with the cross-correlation calculation and the effectiveness of the low-pass filter in reducing the biases of the $A$ and $C$ estimates.
This improvement in seeding will further refine the resolution of the pulse rise time.

\item Enhanced rise-time determination:
Figure~\ref{fig:figure11} elucidates the immediate improvements with the more accurate seeding for the fit parameters on determining the rise time at the pulse level.
The effects are most prominent at lower energies, where the cross-correlation method has manifested the capabilities of correcting the evidently inefficient fits, which, in turn, mitigates misidentified surface and bulk events.
Additionally, the consistency with the existing algorithm at higher energies confirms the reliability of the proposed method.
As a consequence, the bulk-surface discrimination power can be substantially enhanced at higher levels of the analysis.

\item Resolving the bulk-surface convolution:
The accurate determination of the pulse rise time largely inhibits the spectral leakage between bulk and surface events.
With the pulser-generated samples, it has been observed that the magnitude of the spectral leakage in the sub-keV region can be reduced by approximately 70\% with the implementation of the cross-correlation algorithm, as illustrated in Figure~\ref{fig:figure9}.
This is also congruous with the observation of a nearly 70\% decrease on the bulk spectrum in TEXONO reactor data at similar energies, given the dominance of surface leakage at lower energies.
Furthermore, the more confined spread of the bulk and surface spectra allows for better delineation of the event types.
Thus, increasing our ability to resolve the low-energy bulk-surface entanglement at the rise-time distribution level.

\item Lowering analysis threshold:
Finally, with the more constrained rise-time distributions and improved timing-pulse amplitude estimates, the cross-correlation method features substantial enhancement in both the efficiency for bulk selection and the signal retention efficiency for noise cleaning.
Combined with the enhanced ability to discern between bulk and surface events, we have shown in the full energy spectrum analysis using TEXONO data in Section~\ref{sec:texono_data} that this proposed algorithm offers an at least 10~eV$_{ee}$ gain in terms of the analysis threshold.
We consider this the main achievement of the proposed method as this can potentially expand the sensitivity limits and experimental reach of ongoing and future studies targeting for low-energy rare-event searches, e.g., exploring minicharged particles, potential dark matter signatures and coherent elastic neutrino-nucleus scattering, etc.

\item Reduced computation time:
Apart from the obvious achievements in analysis, the proposed cross-correlation approach also exhibits another rather desirable feature, that is, time efficiency.
With the increased precision in parameter estimation and by obtaining the initial estimates for the parameter via a relatively simple calculation instead of a four-parameter fit, the number of fits involved in the new procedure is reduced to 2 from the conventional 4 described in Section~\ref{sec:CC_risetime}.
This will, therefore, reduce the computation time by roughly 50\%.
This advancement in time efficiency can generally help in optimising the overall analysis process, and will be particularly valuable for studies where prompt response is required.

\end{enumerate}

Despite being rather simplistic in design, this proposed algorithm for initial parameter estimation has manifested enormous potential for wider applications and further developments.
The consistency and reliability under tests across a diverse dataset confirms the robustness of the method and states a strong case for broader implementation amongst germanium experiments.
In addition, we believe this algorithmic approach based on pulse-shape methods in obtaining the seeding for fit parameters opens up a new perspective in tackling the bulk-surface discrimination problem.
Future investigations can be built on the current findings to further explore the possibilities of applying more sophisticated pulse-shape techniques in parameter seeding evaluation.
This can very well further advance our abilities in decrypting the information encompassed within the pulses.

We should also note that the software algorithm-based improvements in discrimination power presented in this work should be complemented by hardware advancements on the reduction of the detector pedestal noise level. 
In the mean time, there is still room for improvements in the cross-correlation algorithm when implemented into existing analyses. 
Our next research goal will be its integration and optimisation in the physics analysis with data taken at KSNL. \\

\section*{Acknowledgments}
We are grateful to Dr. Yuki Inoue for the inspiring discussions on data analysis techniques for gravitational wave observatories.
This work is supported by the Academia Sinica Principal Investigator Award No. AS-IA-106-M02, Contracts No. 106-2923-M-001-006-MY5, No. 107-2119-M-001-028-MY3, and No. 110-2112-M-001-029-MY3, from the Ministry of Science and Technology, Taiwan, and 2021/TG2.1 from the National Center of Theoretical Sciences, Taiwan.

\bibliography{references}

\end{document}